\documentclass[fleqn,usenatbib]{mnras}
\usepackage{newtxtext,newtxmath}
\usepackage[T1]{fontenc}
\usepackage{ae,aecompl}
\DeclareRobustCommand{\VAN}[3]{#2}
\let\VANthebibliography\thebibliography
\def\thebibliography{\DeclareRobustCommand{\VAN}[3]{##3}\VANthebibliography}
\usepackage[flushleft]{threeparttable}
\usepackage{graphicx}	
\graphicspath{ {./figures/} }
\usepackage{amsmath}	
\usepackage{tabularx}
\usepackage{makecell, multirow}
\usepackage{gensymb}
\usepackage{pdflscape}
\usepackage[figuresright]{rotating}

\usepackage{upgreek}
\usepackage{float}

\usepackage{booktabs}

\title[Very High Energy FSRQ candidates]{Very High Energy Flat Spectral Radio Quasar Candidates}

\author[Malik Zahoor et al.]{
Malik Zahoor$^{1}$\thanks{E-mail: malikzahoor313@gmail.com},
Sunder Sahayanathan$^{2,3}$\thanks{E-mail: sunder@barc.gov.in},
Shah Zahir$^{4}$\thanks{E-mail: shahzahir4@gmail.com},
Naseer Iqbal$^{1}$,
Aaqib Manzoor$^{1}$
\\
$^{1}$Department of Physics, University of Kashmir, Srinagar 190006, India.\\
$^{2}$Astrophysical Sciences Division, Bhabha Atomic Research Center, Mumbai 400085, India.\\
$^{3}$Homi Bhabha National Institute, Mumbai 400094, India.\\
$^{4}$Department of Physics, Central University of Kashmir, Ganderbal 191201, India.}

\date{Accepted XXX. Received YYY; in original form ZZZ}

\pubyear{2021}

\begin{document}
\label{firstpage}
\pagerange{\pageref{firstpage}--\pageref{lastpage}}
\maketitle

\begin{abstract}
The attenuation of very high energy (VHE) photons by the extragalactic background light (EBL) prevents the observation of
high redshift flat spectrum radio quasars (FSRQs). However, the correlation of VHE spectral index with source redshift
suggests that EBL intensity may be less than what is predicted. This deviation can draw new constraints on opacity of the
universe to VHE gamma-rays. Therefore, more FSRQs may fall above the sensitivity of the forthcoming VHE telescopes than the
ones predicted by the existing EBL models. In order to account for the lower EBL intensity predicted by the index-redshift
correlation, we introduce a redshift dependent correction factor to the opacity estimated from commonly used cosmological
EBL model. Considering this modified opacity, we identify the plausible VHE FSRQ candidates by linearly extrapolating
the \emph{Fermi} gamma-ray spectrum at 10 GeV to VHE regime. Our study suggests among 744 FSRQs reported in \emph{Fermi} fourth catalog-Data
release 2 (4FGL-DR2), 32 FSRQs will be detectable by Cherenkov Telescope Array Observatory (CTAO). Since the FSRQS are proven
to be highly variable, we assume a scenario where the average \emph{Fermi} gamma-ray flux increases by a factor of 10 and this predicts
additional 90 FSRQs that can be detected by CTAO.
\end{abstract}




\section{Introduction}
\label{sec:int}

Very high-energy (VHE, E$>$100GeV) gamma-ray astronomy have the potential to provide unique insights on open issues 
related to cosmology and particle physics. Additionally, it serves as an important probe for multi-wavelength 
and multi-messenger astronomy. VHE emission from the astrophysical sources can be observed from the ground by studying 
the shower of secondary charged particles initiated through the interaction of the primary gamma-rays with 
the atmosphere \citep{2015CRPhy..16..610D}. Among various techniques employed, the Imaging 
Atmospheric Cherenkov Telescopes (IACTs) detect the primary gamma-rays through the image of the Cherenkov pool
caused by the secondary shower.
The current-generation IACTs, which include \emph{HESS} \citep{2006A&A...457..899A}, \emph{MAGIC} \citep{2012APh....35..435A}, and 
\emph{VERITAS} \citep{2006APh....25..391H}, have been contributing extensively to VHE astrophysics for nearly two decades. In tandem with the telescopes operating at other
energies (e.g. \emph{Fermi}-LAT \citep{2009ApJ...697.1071A}, \emph{Swift} \citep{2005SSRv..120..165B}, \emph{NuSTAR} \citep{2013ApJ...770..103H} and \emph{XMM-Newton} \citep{2001A&A...365L...1J}), 
IACTs have provided clues to understand the non-thermal emission processes in blazars \citep{KNODLSEDER2016663}.

The advent of new generation ground-based VHE telescopes, including the proposed Cherenkov Telescope Array Observatory (CTAO) \citep{2019scta.book.....C}, the 
gamma-ray astronomy is entering into a new era. Operating from a few tens of GeV to the multi-TeV energy band, the CTAO is designed to be the 
largest and the most sensitive gamma-ray observatory in this energy range \citep{2021arXiv210804512G}. It will be configured as two sets of Cherenkov telescope arrays, one in each of the Earth's hemispheres, and it is expected to start science operations at full capacity within few years. 
CTAO along with other upcoming VHE experiments, the Large High Altitude Air Shower Observatory (\emph{LHAASO}) \citep{Cao:2021LM} , 
ASTRI Mini-Array (\emph{ASTRI MA}) \citep{Antonelli:2021ml}, Southern Wide-field Gamma-ray Observatory (\emph{SWGO}) \citep{BarresdeAlmeida:2021xgv}, 
Major Atmospheric Cherenkov Experiments (\emph{MACE}) \citep{2021arXiv210704297H} etc, will be able to explore the gamma-ray sky with unprecedented performance, notably in the 
multi-TeV energy range.

Blazars dominate the extragalactic sky at VHE energies. These objects are the subclass of Radio loud active galactic nuclei (AGNs) having a
relativistic jet pointing close to the line of sight of the observer on earth \citep{Urry_1995}.
The spectral energy distribution (SED) of blazars is dominated by the non thermal emission from the jet and consists of two broad peaks.
The low energy component extends from radio-to-X-rays and is attributed to the synchrotron emission, while the high energy component is commonly 
interpreted as the inverse Compton (IC) scattering of low energy photons by the relativistic electron distribution in the jet \citep{Urry_1995}. 
Blazars are further classified into BL Lac objects and Flat spectrum radio Quasars (FSRQs) based on the presence/absence of broad emission/absorption 
line features in their optical spectra. 
The synchrotron SED of BL Lacs generally peaks at optical-to-X-ray energies; whereas, for FSRQs this spectral peak 
fall in infrared-optical energy range. Besides the variation in peak location, the IC component of blazars is significantly different 
for BL Lacs and FSRQs. Particularly, the IC luminosity for FSRQs is larger than the synchrotron luminosity, commonly referred as 
    {\it Compton dominance}, while it is of similar order in case of BL Lacs \citep{1994ApJ...421..153S}. The target photon field for the IC process is
also different for these two class of blazars. The IC scattering of synchrotron photons (synchrotron self Compton mechanism: SSC) 
is quite successful in explaining 
the Compton spectral component of BL Lacs. However, modelling the high energy SED of FSRQs through IC mechanism demands additional 
photon field other than synchrotron emission. This photon field can be external to the jet and the plausible sources are
the thermal photons from the accretion disc \citep{1993ApJ...416..458D}, broad emission line photons \citep{B_a_ejowski_2000} and/or the IR photons from the dusty torus \citep{1994ApJ...421..153S}.

The interaction of VHE gamma-rays with the extragalactic background light (EBL) results in the formation of electron/positron pairs \citep{DWEK2013112}. The amount of attenuation depends on the redshift of the source and the energy of the VHE photons. For distant blazars, this causes the observed VHE spectra to steepen, causing the flux to fall below the telescope sensitivity. Consequently this makes the Universe opaque above few tens of GeV for objects having larger redshifts (gamma-ray horizon) which was initially predicted at 
z > 0.1 \citep{1967PhRv..155.1404G}. Improved sensitivity of the current generation Cherenkov telescopes and better estimation 
of EBL intensity have significantly modified the gamma-ray horizon; nevertheless, detection of the FSRQs 4FGL\,J0221.1+3556 and 4FGL J1443.9+2501 located at 
redshifts 0.954 \citep{2016A&A...595A..98A} and 0.939 \citep{2015ApJ...815L..22A, 2015ApJ...815L..23A} suggest the
EBL intensity may still be less than the prediction.

Knowledge of EBL intensity and the intrinsic source VHE spectrum is crucial in identifying the VHE blazar candidates. The latter is often
estimated by extrapolating the \emph{Fermi} high energy gamma-ray spectrum to VHE energies \citep{2021MNRAS.508.6128P, 2021ApJ...916...93Z}. The VHE blazar candidates are then obtained 
by convolving the extrapolated spectrum with EBL induced opacity predicted through cosmological models. However, such candidates are put forth
only for BL Lacs \citep{Massaro_2013,10.1093/mnras/stz812,10.1093/mnras/stz3532,10.1093/mnras/stz3018,2021MNRAS.508.6128P,2021ApJ...916...93Z} and no such study have been performed for FSRQs. The primary reason being, the \emph{Fermi} spectrum of FSRQs generally
deviate from a power-law and is often represented by a log-parabolic function, and extending this function to VHE energies roll off the spectrum significantly.
Additionally, the FSRQs are populated at higher redshifts and the current EBL models disfavour them as probable candidates. Besides this, most VHE detections of FSRQs have been during enhanced flux states and hence the flux variability plays a crucial role in VHE studies of these sources.

 In our earlier study \citep{2022MNRAS.511..994M} (hereafter Paper I), using the correlation between observed VHE spectral index with the redshift, we have highlighted the deviation of different cosmological EBL models from the observations. The important assumptions in the earlier study are: a) The average intrinsic VHE spectral index is consistent with the regression line extrapolated to redshift, z=0, b) The spectral index variation of the individual source is much smaller than the steepening introduced by EBL at large redshifts, and c) The cosmological evolution of the source do not modify the intrinsic VHE spectral indices. This study suggests the EBL intensity may be much less intense than the ones predicted by the cosmological models.
Consistently, this also suggests the gamma-ray horizon may fall at much larger redshifts than the one presumed. Moreover, from the 
X-ray spectral studies of blazars it is known that the log-parabolic function is successful in reproducing only a narrow energy 
band \citep{2004A&A...413..489M}. Hence, it may not be judicious to expect the \emph{Fermi} log-parabolic spectral shape to extend up to VHE.
In this work, we predict the plausible VHE FSRQ candidates considering these discrepancies.  We perform a linear extrapolation
of the high energy spectrum of FSRQs listed in 4FGL-DR2 catalog \citep{2020arXiv200511208B} to VHE as a prediction for the intrinsic VHE spectra.
To account for the reduction in EBL intensity, suggested by VHE observations of FSRQs (Paper I), we add a redshift dependent 
correction factor to the EBL opacity provided by \cite{Franceschini_2017} (hereafter FM). These modifications are then used to predict the 
list of VHE FSRQs that can be studied by CTAO and other operational Cherenkov telescopes. In the next section \S \ref{sec:fsrq_cand} we first introduce a correction factor to the existing EBL estimates (using Paper I) followed by intrinsic VHE flux estimations using \emph{Fermi} spectral information. The section concludes by over-plotting the sensitivity curves of present and upcoming VHE telescopes to look for the possible VHE FSRQ candidates. Finally the Results are summarised and discussed in section \S \ref{sec:discussion}. In this work we adopt a cosmology with $\rm \Omega_M = 0.3$, $\rm \Omega_\Lambda = 0.7$, and $\rm H_0 = 71  km s^{-1} Mpc^{-1}$.

\section{VHE FSRQ Candidates}
\label{sec:fsrq_cand}

\subsection{Modified EBL Photon Density}
\begin{figure}
		\centering
		\includegraphics[scale=0.3, angle=270]{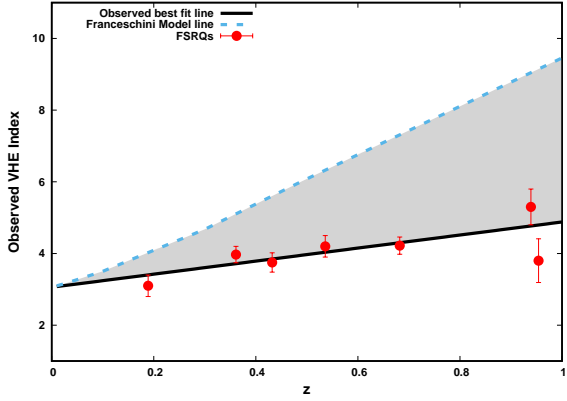}
		\caption{Comparison of observed VHE spectral indices with those estimated using EBL model by \citet{Franceschini_2017} (FM) for FSRQs. The black solid line represents the best fit straight line to 7 FSRQs detected in VHE (Paper I), dashed line represents the index values estimated using FM. Grey region shows the deviation of EBL model from the observational best fit line to VHE indices.} 
		\label{fig:fig1}
\end{figure}

The VHE detection of large redshift FSRQs indicate the Universe may be more transparent to VHE gamma-rays than anticipated. A similar conclusion is also drawn from the correlation study between the observed  
VHE spectral index and the redshift (Paper I). These results suggest, the predicted EBL intensity at 
large redshifts may be considerably larger than the actual value. To account for this excess, we introduce
a redshift dependent correction factor $a(z)$ to the commonly used EBL opacity $\tau(E,z)$ by FM,
\begin{align}\label{eq:taucorr}
\tau_{c}(E,z) = a(z) \tau(E,z)
\end{align}
where, $\tau_c$ is the corrected opacity (hereafter modified FM; MFM) and $E$ is the VHE photon energy. This correction factor will result in a  
harder observed VHE spectral index $\Gamma_{\rm obs}$ for a putative intrinsic spectrum (Figure~\ref{fig:fig1}) given by Paper I,
\begin{align}
	\Gamma_{\rm obs} = \Gamma_{\rm int} + a(z) \frac{d\,\tau}{d\,ln(E)}
\end{align}
where, $\Gamma_{\rm int}$ is the intrinsic VHE spectral index and $a(z) < 1$. Estimation of $a(z)$ demands the 
knowledge of $\Gamma_{\rm int}$ which is obtained from the y-intercept of the best fit straight line to 
$\Gamma_{\rm obs}$ and $z$ distribution (Paper I).

\subsection{Intrinsic VHE Source Flux}
The opacity of VHE photons, along with the correction factor, equation (\ref{eq:taucorr}), let us to derive the
observed flux of FSRQs provided the intrinsic flux is known. To estimate the latter we perform a linear 
extrapolation of the high energy spectrum of FSRQs to VHE. The information regarding the high energy spectrum 
is obtained from the fourth \emph{Fermi} catalog, 4FGL-DR2 \citep{2020arXiv200511208B}. The data used in this catalog was collected 
over the period of ten years, starting from August 4 (15:43 UTC) in 2008 to August 2 in 2018 (19:13 UTC). 
It employs the same analysis methods as of 4FGL catalog \citep{2020ApJS..247...33A} in the energy range of 50 MeV to 1 TeV. 
Among 5788 sources listed in the catalog, 744 fall into the list of FSRQs. The attenuation due to EBL 
depends upon the source redshift and this is obtained from the online databases 
NED\footnote{\url{https://ned.ipac.caltech.edu/}} and SIMBAD\footnote{\url{http://simbad.cfa.harvard.edu/simbad/}}. 
These databases provide the redshifts of 586 FSRQs listed in 4FGL-DR2 and hence only these sources are considered for the 
present study. The information regarding the high energy spectrum of these FSRQs is extracted from the xml file available 
online\footnote{\url{https://fermi.gsfc.nasa.gov/ssc/data/access/lat/10yr_catalog/gll_psc_v26.xml}} and 
the same is plotted as a black dotted line in Figure~\ref{fig:fig2}.

The \emph{Fermi} spectral behaviour for the selected sample of FSRQs is restricted to three distinct shapes. Among the 586 
individual spectra of the sources, 263 are logparabola, 321 power-law and 2 sources exhibit a power-law with exponential cutoff. For the extrapolation to VHE band, we assume a power-law extension with index consistent with the spectral slope
at 10 GeV (See Table~\ref{tab:tab1} \& \ref{tab:tab2}). This energy is chosen since attenuation due to EBL is minimal below this energy for the range of redshifts considered \citep{2012Sci...338.1190A}, and the \emph{Fermi} spectrum can be treated as true intrinsic spectrum. For 8 sources, we find the terminal \emph{Fermi} power-law index is harder ($\lesssim$ 2). Such hard
indices suggest the Compton peak of these sources lie beyond the \emph{Fermi} energy range and the VHE spectra can be much steeper than the one predicted by the power-law extrapolation. Hence we exclude these sources from this study.
The extrapolated power law from the \emph{Fermi} fit to the VHE band for the rest of sources is shown in Figure~\ref{fig:fig2} 
as a red dotted line and this serves as the intrinsic VHE spectra $F_{\rm int}$ for the present study. 
The observed VHE spectrum $F_{\rm obs}$ is then obtained using $\tau_c$ from equation (\ref{eq:taucorr}) (MFM)
\begin{align}
	F_{\rm obs} (E,z) = F_{\rm int}(E)\,e^{-\tau_c(E,z)}
\end{align}
The EBL attenuated VHE spectrum for the selected FSRQs using MFM is shown as 
solid black line in Figure~\ref{fig:fig2}. The attenuated spectrum considering FM is shown as green dotted line.

\subsection{Comparison with IACT sensitivity}
To identify the FSRQ candidates detectable by the operational and upcoming IACTs, we compare the predicted EBL attenuated
VHE spectrum with the available sensitivity curves. 
The sensitivity curves of \emph{CTAO}-South and North (Zenith Angle, ZA $<$ 20 deg) are obtained from the \emph{CTAO} 
webpage\footnote{\url{https://www.cta-observatory.org/science/cta-performance/}}
while the sensitivity curve for \emph{VERITAS} (ZA $<$ 40 deg)
is obtained from \emph{VERITAS} 
webpage\footnote{\url{https://veritas.sao.arizona.edu/about-veritas/veritas-specifications/}}. 
For other instruments such as \emph{MAGIC} (ZA $<$ 35 deg) and \emph{H.E.S.S} (ZA $<$ 18 deg), the sensitivity curves are obtained from 
\citet{2016APh....72...76A} and \citet{Holler:2016Av} respectively. These sensitivities are calculated at 5$\sigma$ significance for 50 hours exposure time and are shown in Figure~\ref{fig:fig2} with legends.

Comparing sensitivity of the IACTs and the predicted VHE spectrum of \emph{Fermi} detected FSRQs, we identify the plausible
VHE FSRQ candidates and list them in Table~\ref{tab:tab1}. Our study suggest, 32 FSRQs would fall within the detection
threshold of \emph{CTAO's} Omega configuration (full-scope configuration), while the detection status with \emph{CTAO}'s first construction phase (Alpha configuration) will be 29 using opacity estimates from MFM. Considering FM, the number of sources falling under the detection status of CTAO reduce to 23 and 20 for Omega and Alpha configurations of \emph{CTAO} respectively. 
With the operational IACTs, the number of FSRQs falling within the detection threshold are 5 for \emph{VERITAS} and 2 sources each for \emph{MAGIC} and \emph{HESS} using opacity estimates from MFM, while using FM the number of sources reduce to 2, 1 and 1 for \emph{VERITAS}, \emph{MAGIC} and \emph{HESS} respectively (Table~\ref{tab:tab1}).

Blazars being extremely variable \citep{2010ApJ...722..520A, 2019ApJ...877...39M, 2020A&A...634A..80R}, the intrinsic VHE flux derived from cumulative \emph{Fermi} observations
may portray only the average spectrum. To account for this variability, we assume a scenario where the VHE flux enhances by a factor of 10 above the prediction.
Further, to be consistent with the rapid variability we use the \emph{CTAO} sensitivity corresponding to 5 hour exposure time only. We refer the FSRQs which fall within the detection threshold of \emph{CTAO} under this assumption as marginally detectable. In Table~\ref{tab:tab2}, we list the FSRQs which are marginally detectable and find additional 82 sources can be detected under Aplha configuration of \emph{CTAO} for the opacity estimated by MFM. This number increases to 90 under Omega configuration of CTAO. For the opacity estimated from FM, the number of sources falling under the marginal detection list are 43 and 40 with Omega and Alpha configurations of \emph{CTAO} respectively. 

\section{Summary \& Discussion}
\label{sec:discussion}
The IACTs are narrow angle telescopes and require long duration observation of distant faint sources for significant detection. 
Hence, identification of plausible candidate FSRQs can provide a guideline for the upcoming VHE telescopes.
Our earlier study, based on correlation between observed VHE spectral index with redshift (Paper I) suggests the universe may 
be more transparent to VHE photons than those predicted by cosmological EBL models. Taking cue from this, we 
predict the observed VHE fluxes of \emph{Fermi} detected FSRQs and compare the same with the sensitivity of operational/upcoming
telescopes. We find a significant number of FSRQs, listed in Table~\ref{tab:tab1} and Table~\ref{tab:tab2}, can be studied by CTAO while few of them even
with the operational IACTs. The sources which fall within the sensitivity limits of operational IACTs and are not reported in VHE yet are 4FGL\,J0043.8+3425 (\emph{VERITAS}) and 4FGL\,J0957.6+5523 (\emph{VERITAS}, \emph{MAGIC}) (see Table~\ref{tab:tab1}). 
Among them 4FGL\,J0043.8+3425, is located at $z=0.966$ which will be the second distant FSRQ if detected and only next to the newly announced 4FGL\,J0348.5-2749
at z = 0.991 \citep{2021ATel15020....1W}. However, the 10 GeV spectral index for this source is $\sim$2.08. Though this satisfies our selection criteria,
the hard index suggests the Compton spectral peak may
fall close to 10 GeV and hence the extrapolation can be questionable. Interestingly, the VHE emission from the FSRQ 4FGL\,J1159.5+2914, newly announced by \emph{VERITAS} \citep{2017ATel11075....1M} and \emph{MAGIC} \citep{2017ATel11061....1M} is also 
predicted to be a candidate source in this work (see Table~\ref{tab:tab1}). We also examined the detection status for \emph{SWGO}, \emph{LHAASO} and \emph{MACE}, however we did not find any source falling within the detection threshold of these instruments.

Among the FSRQs predicted for CTAO, we find 11 sources at $z>1$ fall under the detectable list and 33 under marginally detectable list (see Table~\ref{tab:tab1},\ref{tab:tab2}).
If detected, these high redshift sources may pose challenges to the existing cosmological EBL models. Alternatively, it can
also play an important role in understanding the cosmology. Limits on EBL is mainly obtained through numerical models of the 
galaxy formation and/or their evolution with appropriate cosmological initial conditions \citep{2011MNRAS.410.2556D}. The model parameters 
are fine tuned to reproduce the observed universe. VHE identification of the sources at large redshifts can therefore be an 
important element to constrain 
the EBL which in turn can provide a better understanding about the galaxy formation and evolution. These identifications can also be used to test the 
alternate theories involving oscillation of photons and axion like particles proposed by the standard model \citep{2018PrPNP.102...89I,2018JHEAp..20....1G}.

Considering the fact that the blazars are extremely variable \citep{2010ApJ...722..520A, 2019ApJ...877...39M, 2020A&A...634A..80R}, certain FSRQs may still be 
detectable by operational/upcoming telescopes even though our study suggests otherwise. For instance, the FSRQ 4FGL\,J1422.3+3223 falls below our criteria for detection though it is detected at VHE \citep{2021A&A...647A.163M}. This probably indicates 
that during the flaring epoch, the flux of this source can enhance more than 10 of its average flux. Consistently, 
we find the \emph{Fermi} flux of 4FGL\,J1422.3+3223 during the VHE detection is typically more than 100 times larger than the average flux quoted in 4FGL catalog \citep{2020ATel13382....1C}. Considering a similar factor of flux enhancement in the present work would make 
this source also as detectable by the operational IACTs \emph{MAGIC} and \emph{VERITAS}. Similar conclusion can be arrived for the newly
announced FSRQ 4FGL\,J0348.5-2749 where the increase in Fermi flux, contemporaneous to the VHE detection, was $\sim$200 times 
compared to average flux reported in \emph{Fermi} 4FGL catalog \citep{2021ATel15020....1W}.

The VHE FSRQ candidates predicted in this work depend on 
the choice of $a(z)$ and the robustness of the regression line.
However, the regression line is obtained by fitting merely 7 data points and this may deviate with future detections. Since the 
estimation of $a(z)$ assumes the intrinsic VHE index to be the y-intercept  of the regression line, any deviation in the fit parameters
can modify these predictions considerably. Conversely, a better regression analysis needs more FSRQs to be detected in VHE 
and the prediction based on the available information can facilitate this requirement. Detection of more VHE FSRQs with precise
index measurements will also let us fit the redshift-index dependence with non-linear functions. Such a study will also have a 
major role in constraining the cosmological models. 

VHE spectrum of FSRQs is better explained by a power-law function and hence we have assumed the intrinsic  source 
spectrum also to be a power-law. If we consider the EC interpretation for the VHE emission, the Klein-Nishina effects 
will be substantial and the spectrum will deviate from a simple power-law \citep{1993ApJ...416..458D}. In addition, emission
at VHE may involve high energy electrons that may fall close to the cut-off energy of the underlying electron 
distribution \citep{1998A&A...333..452K}. Under these conditions, the intrinsic VHE spectrum may deviate
significantly from a powerlaw and hence the powerlaw extrapolation to VHE energies can be an overestimate. Therefore these predictions on stringent conditions should be treated as an upper limit. The Klein-Nishina effect will depend upon the energy of the target photons and under extreme limits
the VHE spectrum will be very steep with the photon spectrum nearly following the electron distribution \citep{1970RvMP...42..237B}.
Hence, VHE study of FSRQs can also have the potential to understand the photon field environment of FSRQs.

\section{Acknowledgement}
The authors thank the anonymous referee for valuable comments and suggestions. M.Z, S.S, N.I \& A.M acknowledge the financial support provided by Department of Atomic energy (DAE), Board of Research in Nuclear Sciences (BRNS), Govt of India via Sanction Ref No.: 58/14/21/2019-BRNS. SZ is supported by the Department of Science and Technology, Govt. of India, under the INSPIRE Faculty grant (DST/INSPIRE/04/2020/002319). This research has made use of the CTA instrument response functions provided by the CTA Consortium and Observatory, see https://www.cta-observatory.org/science/cta-performance/ (version prod3b-v2; https://doi.org/10.5281/zenodo.5163273 and version prod5 v0.1; https://doi.org/10.5281/zenodo.5499840) for more details.

\section{Data Availability}
The codes and model used in this work will be shared on a reasonable request to to the corresponding author Malik Zahoor (email:
malikzahoor313@gmail.com).

\newpage
\begin{figure*}
\includegraphics[width=0.30\textwidth]{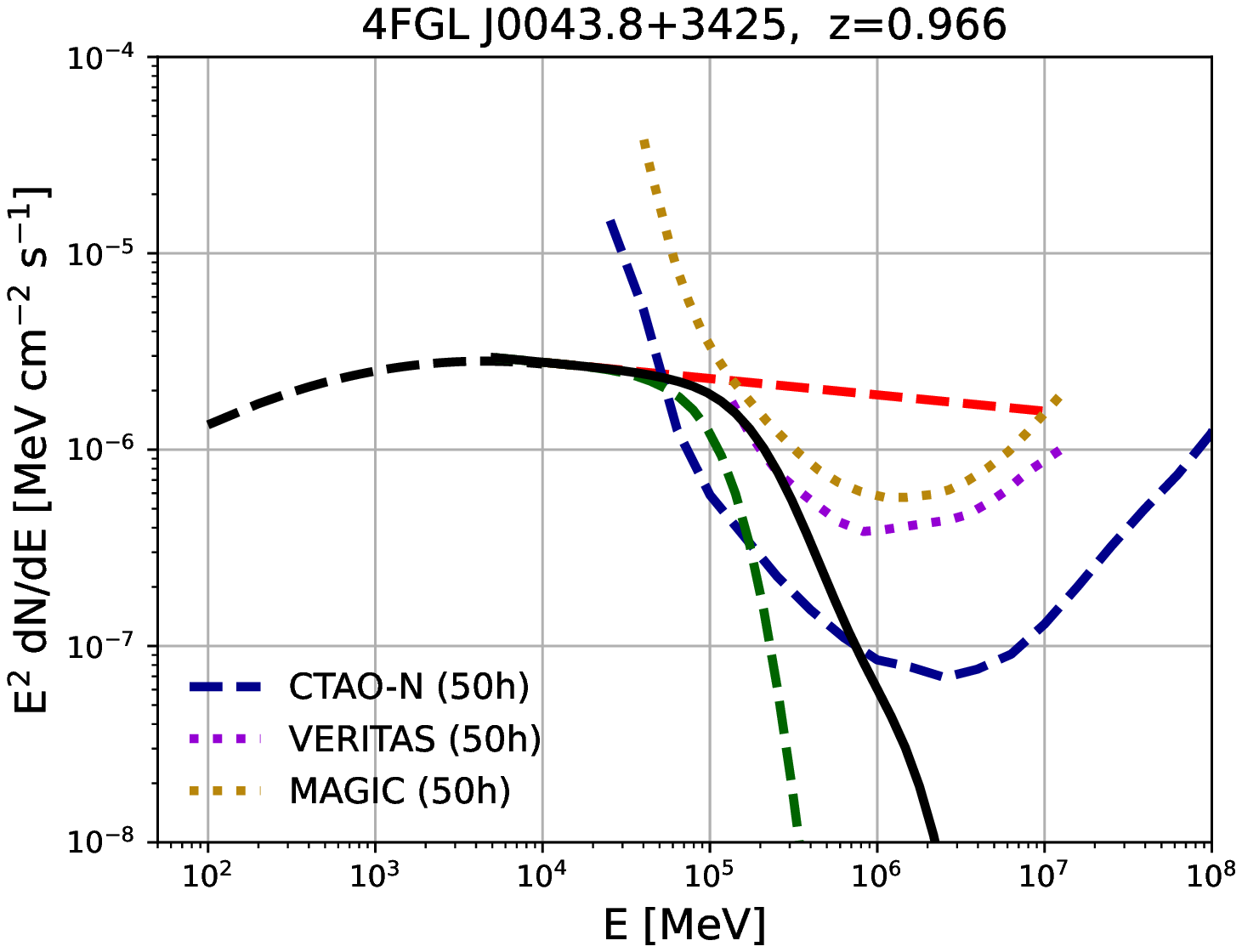}
\includegraphics[width=0.30\textwidth]{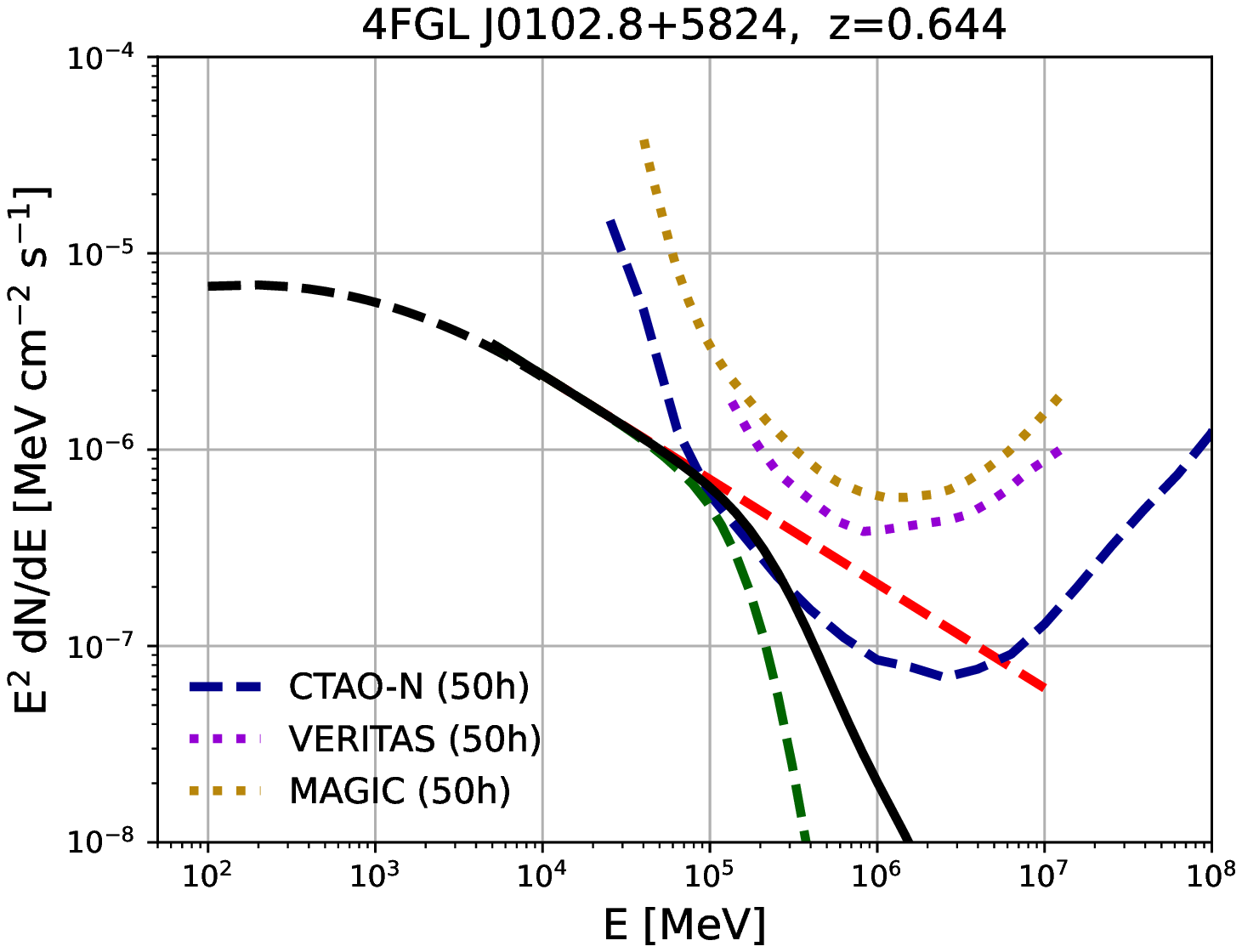}
\includegraphics[width=0.30\textwidth]{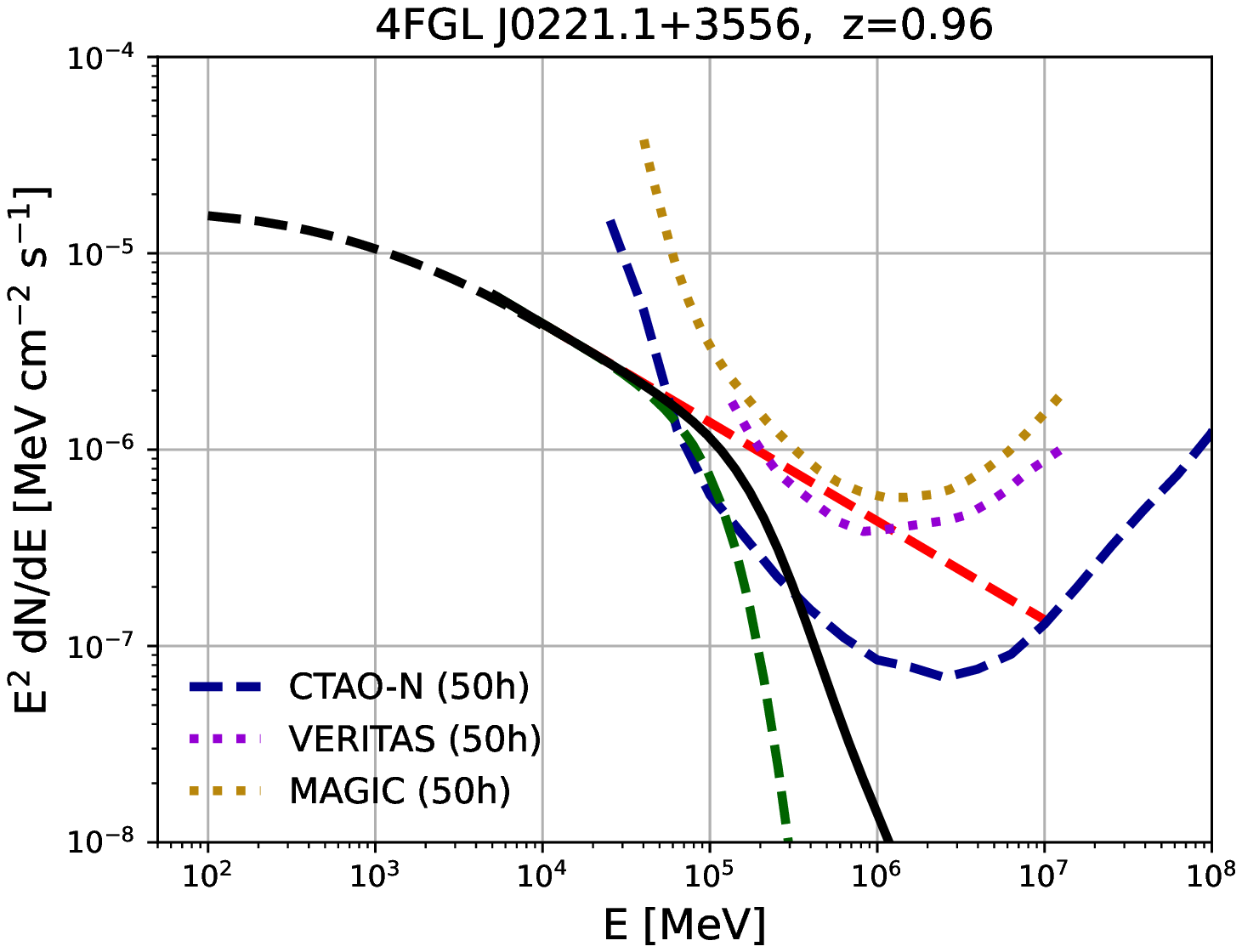}
\includegraphics[width=0.30\textwidth]{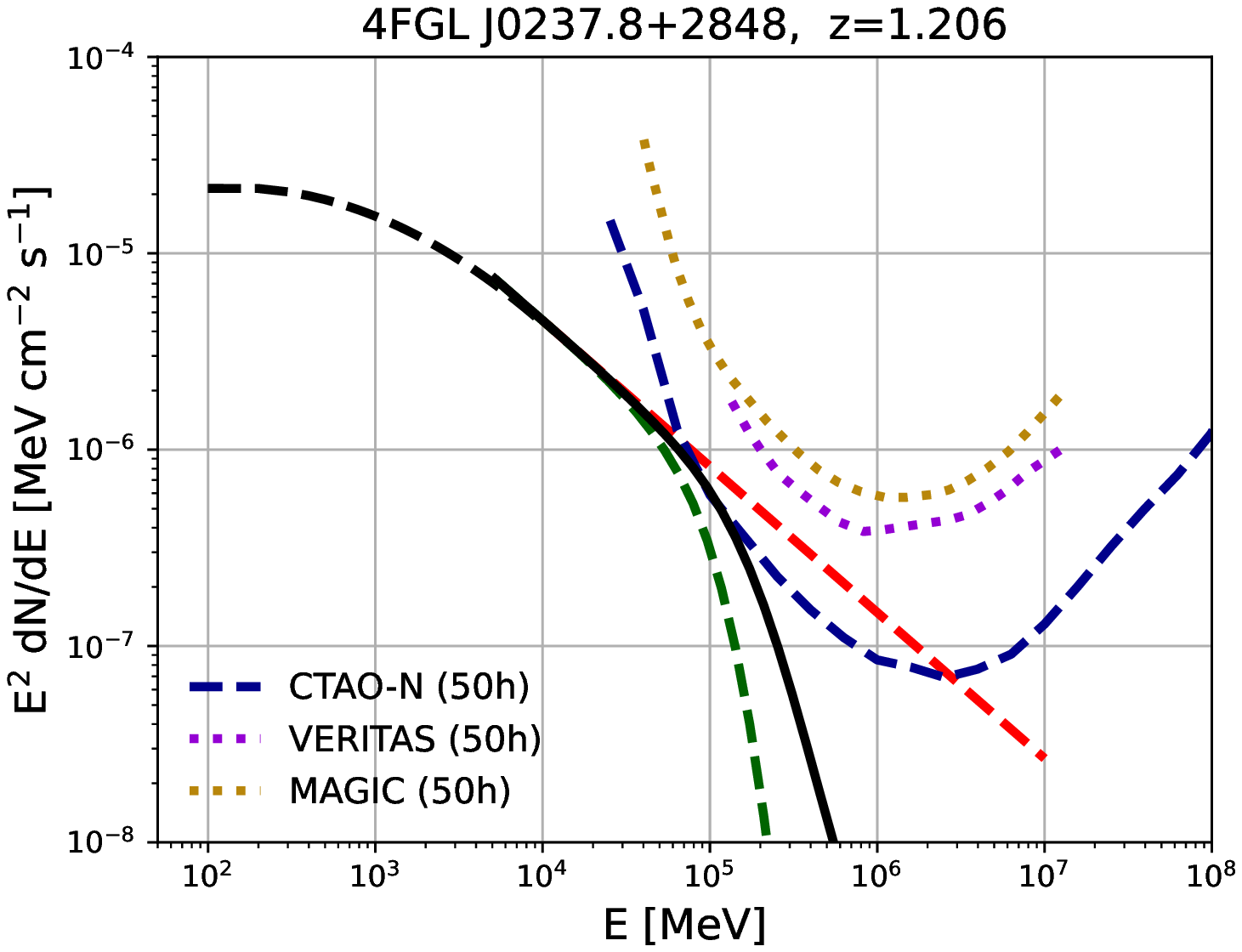}
\includegraphics[width=0.30\textwidth]{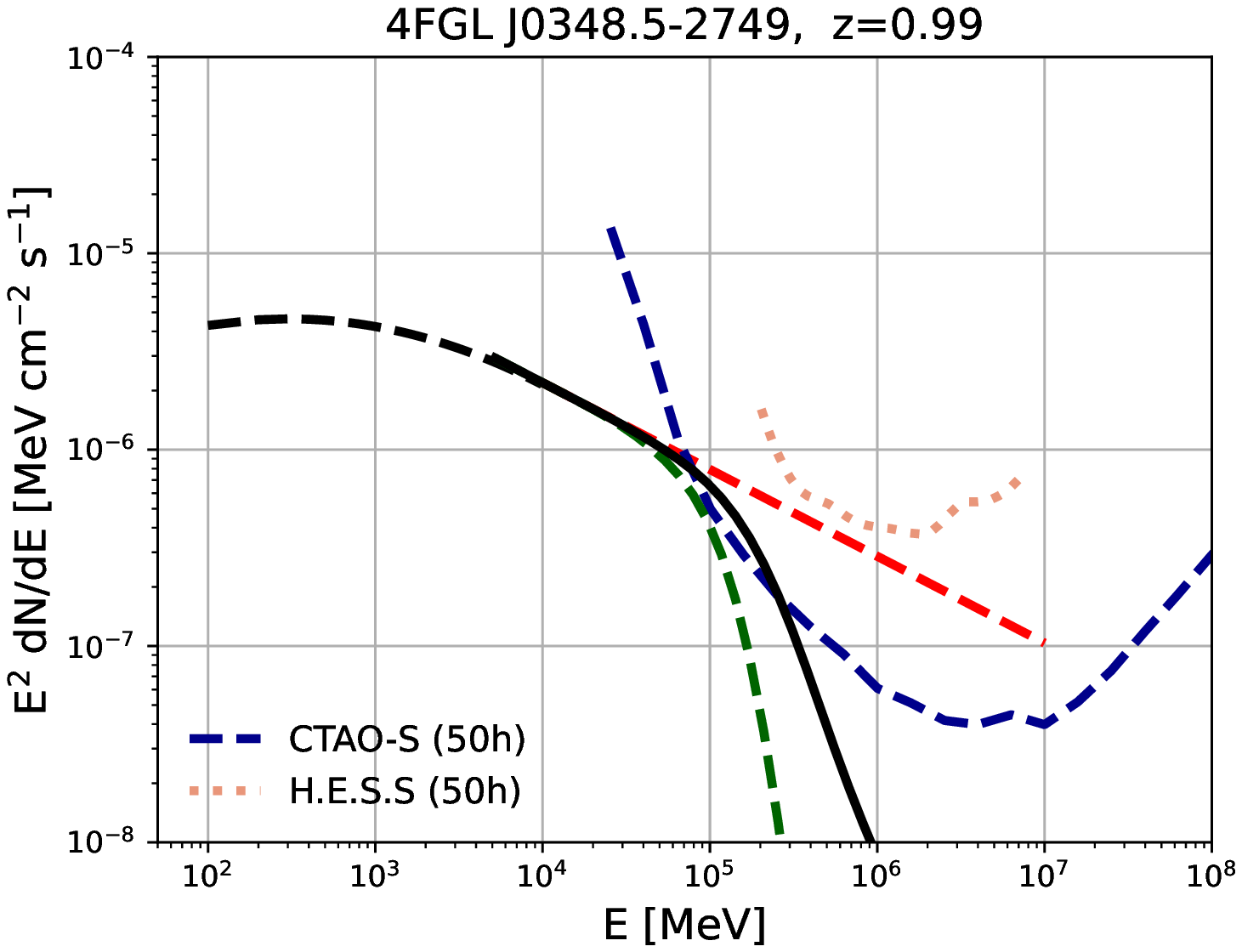}
\includegraphics[width=0.30\textwidth]{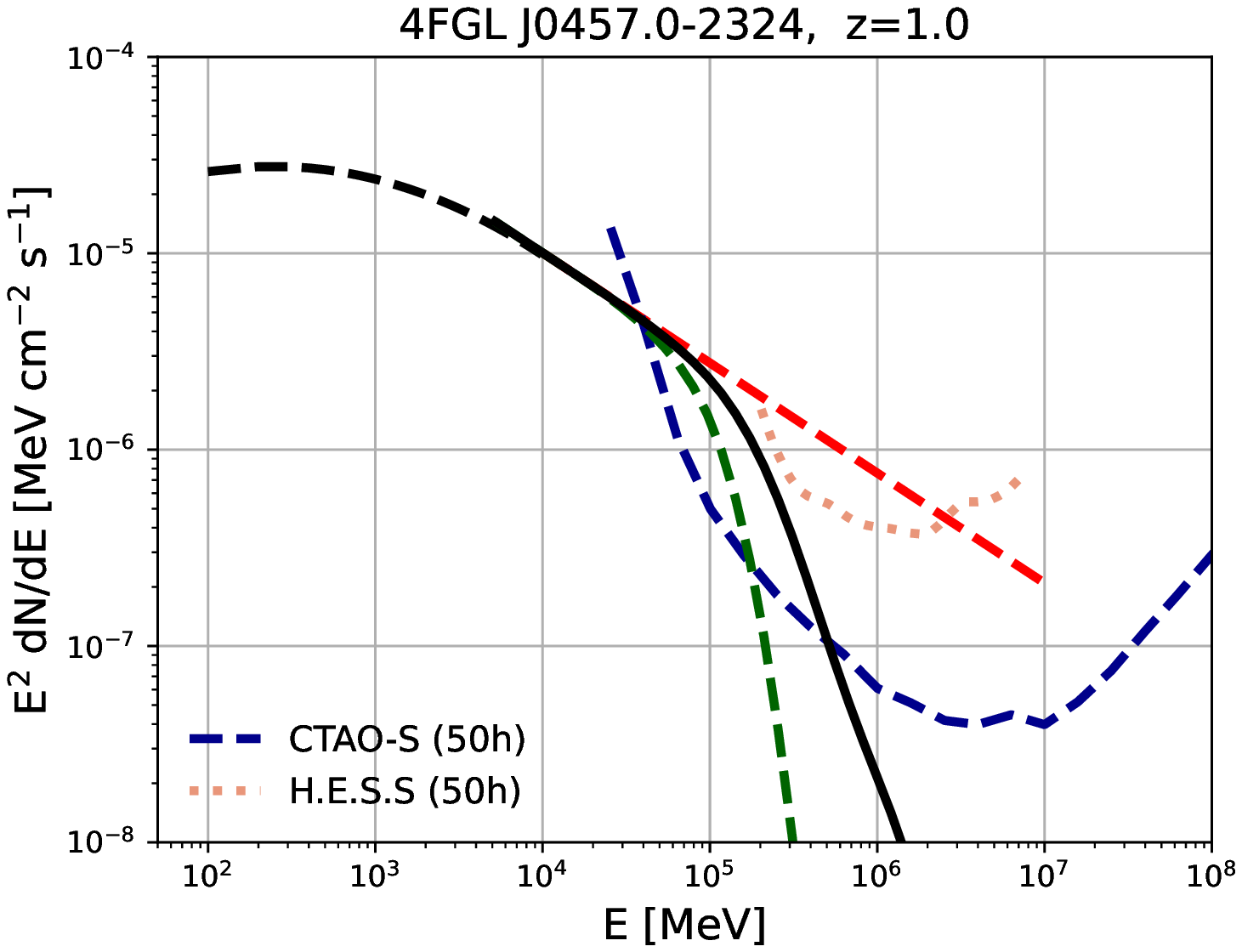}
\includegraphics[width=0.30\textwidth]{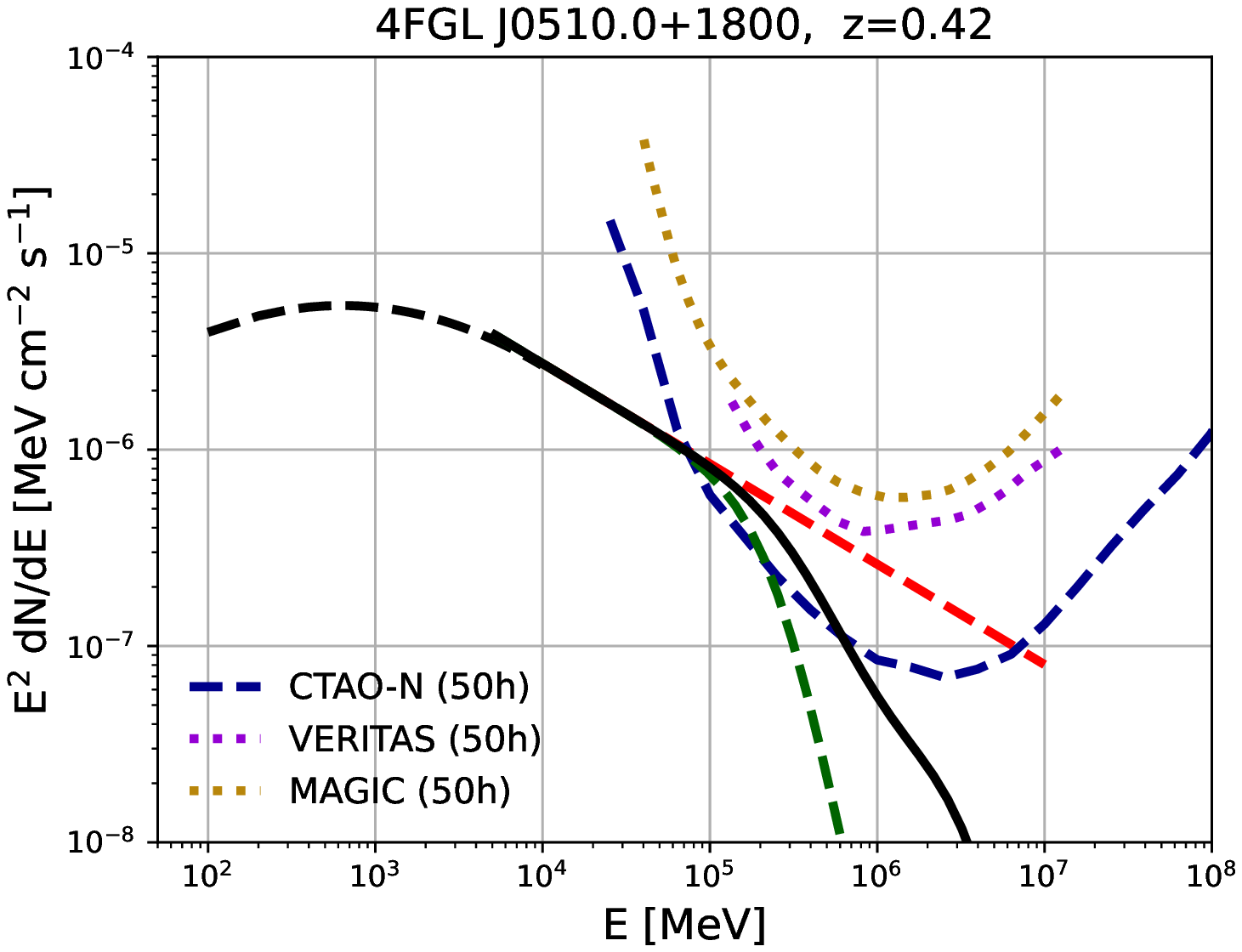}
\includegraphics[width=0.30\textwidth]{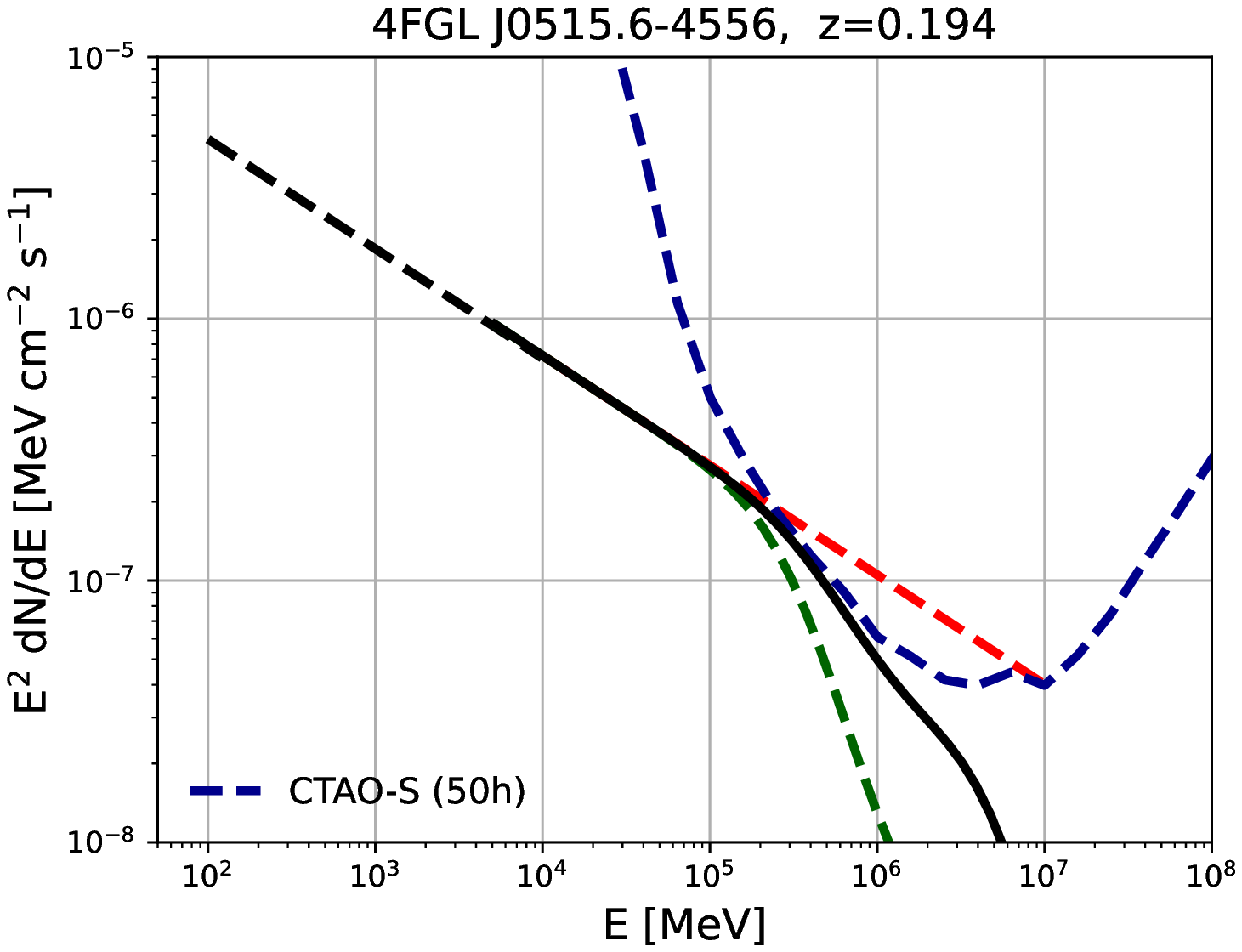}
\includegraphics[width=0.30\textwidth]{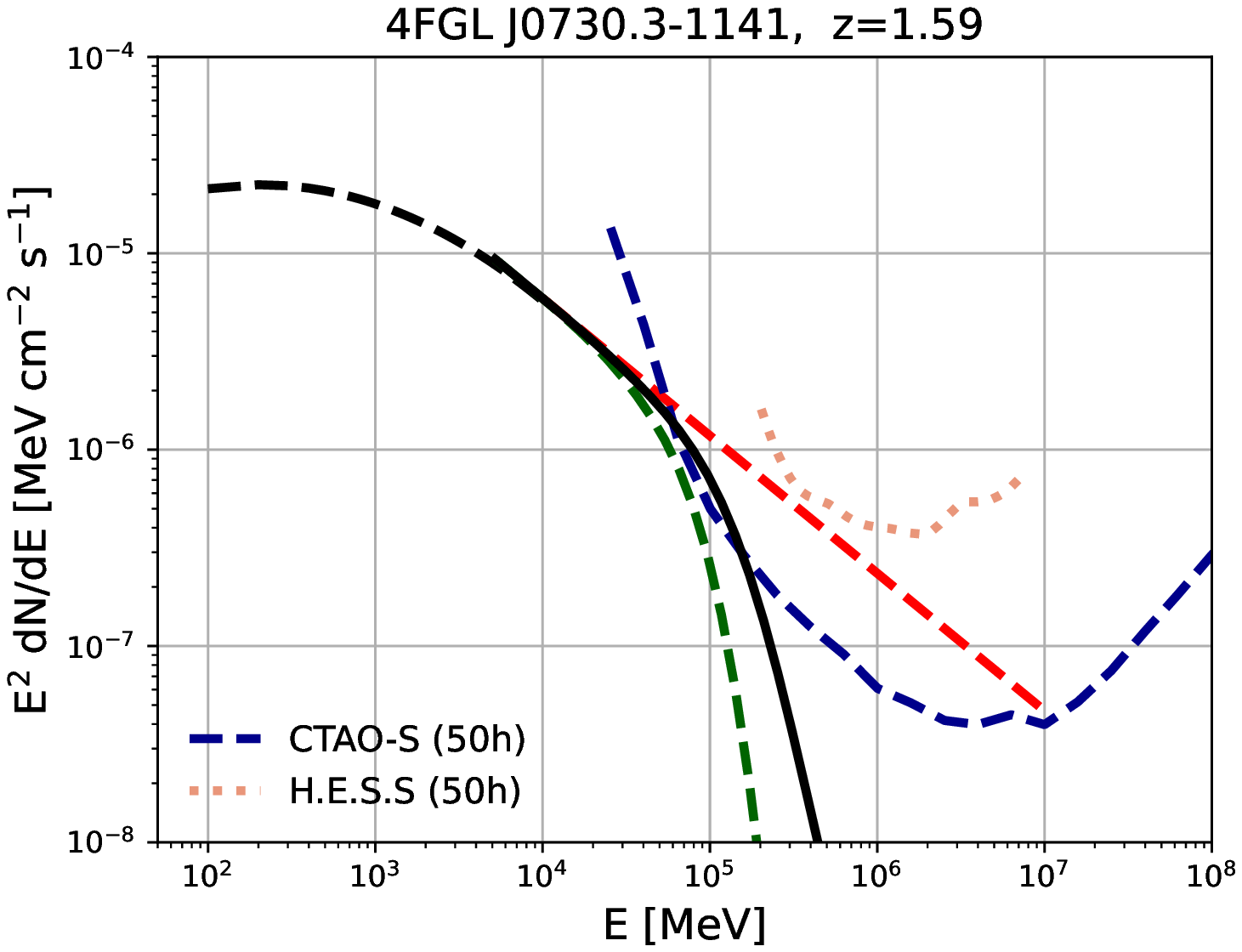}
\includegraphics[width=0.30\textwidth]{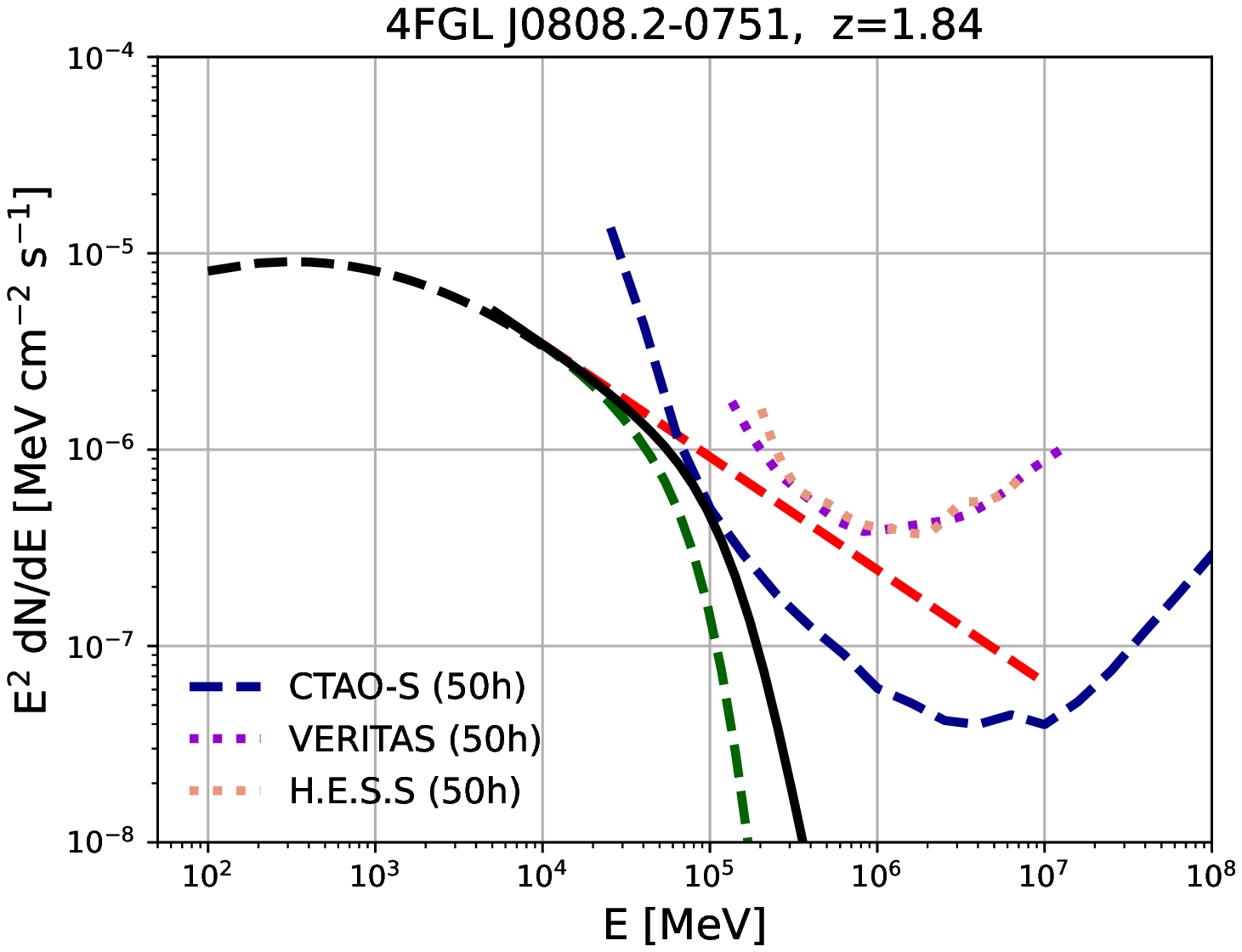}
\includegraphics[width=0.30\textwidth]{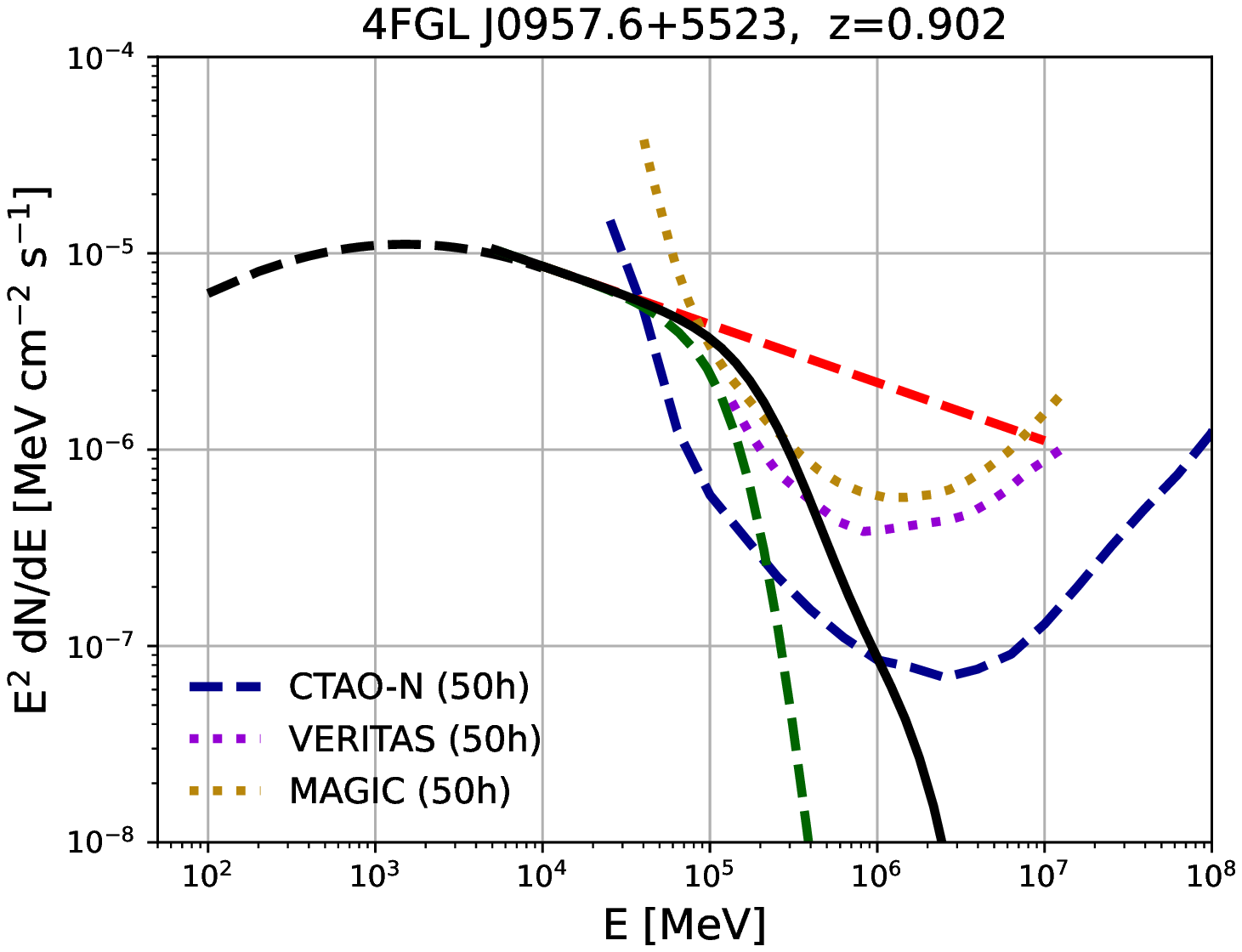}
\includegraphics[width=0.30\textwidth]{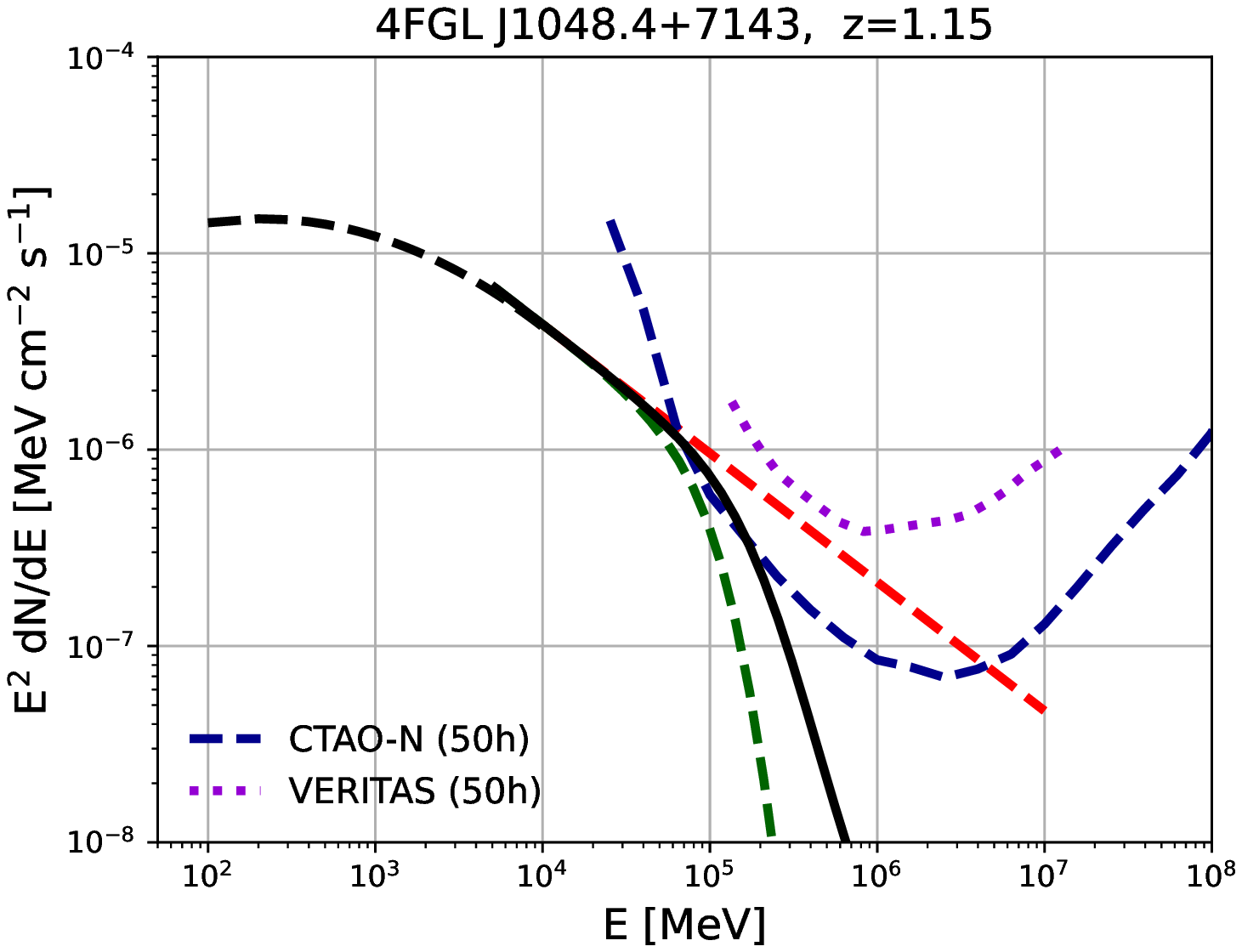}
\includegraphics[width=0.30\textwidth]{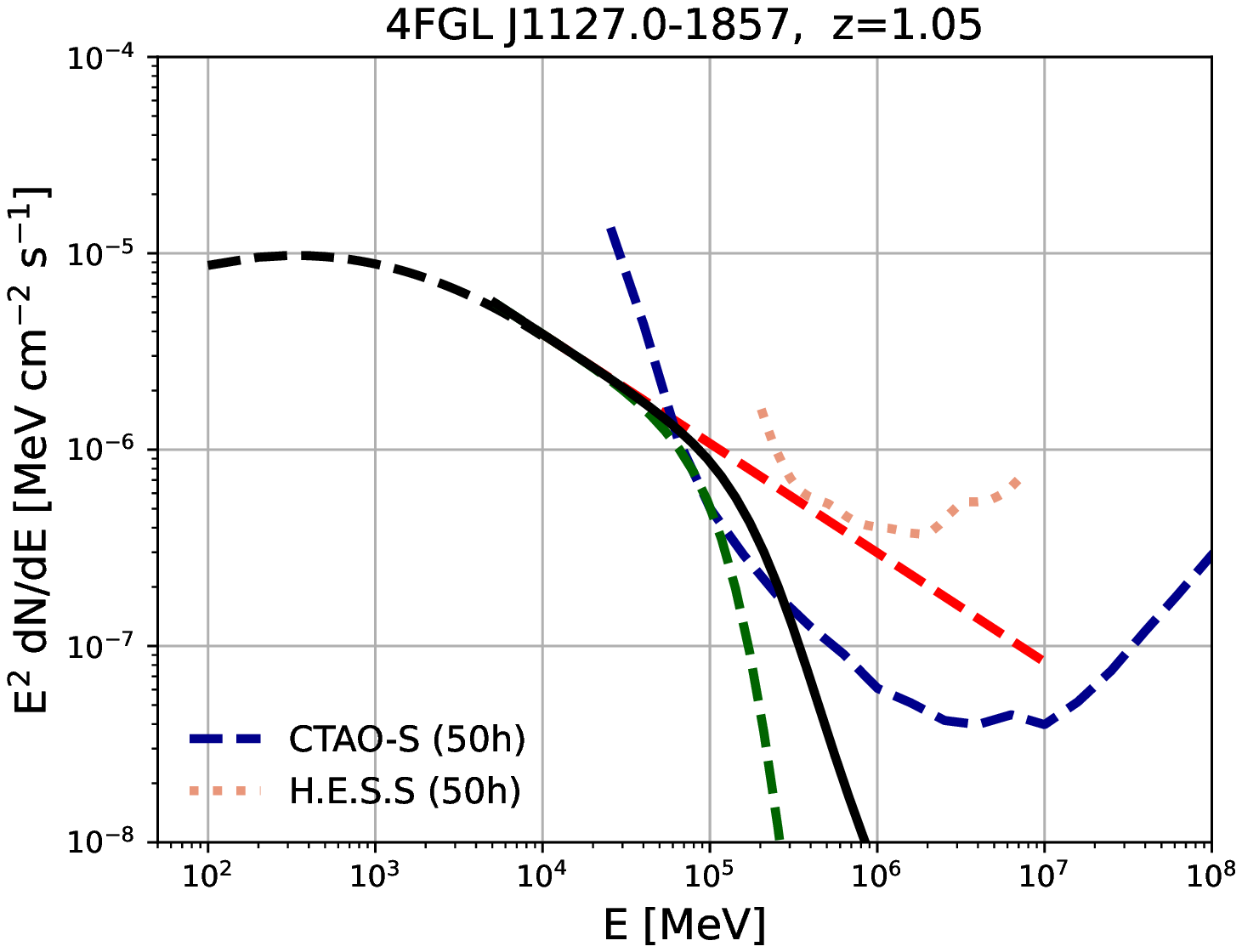}
\includegraphics[width=0.30\textwidth]{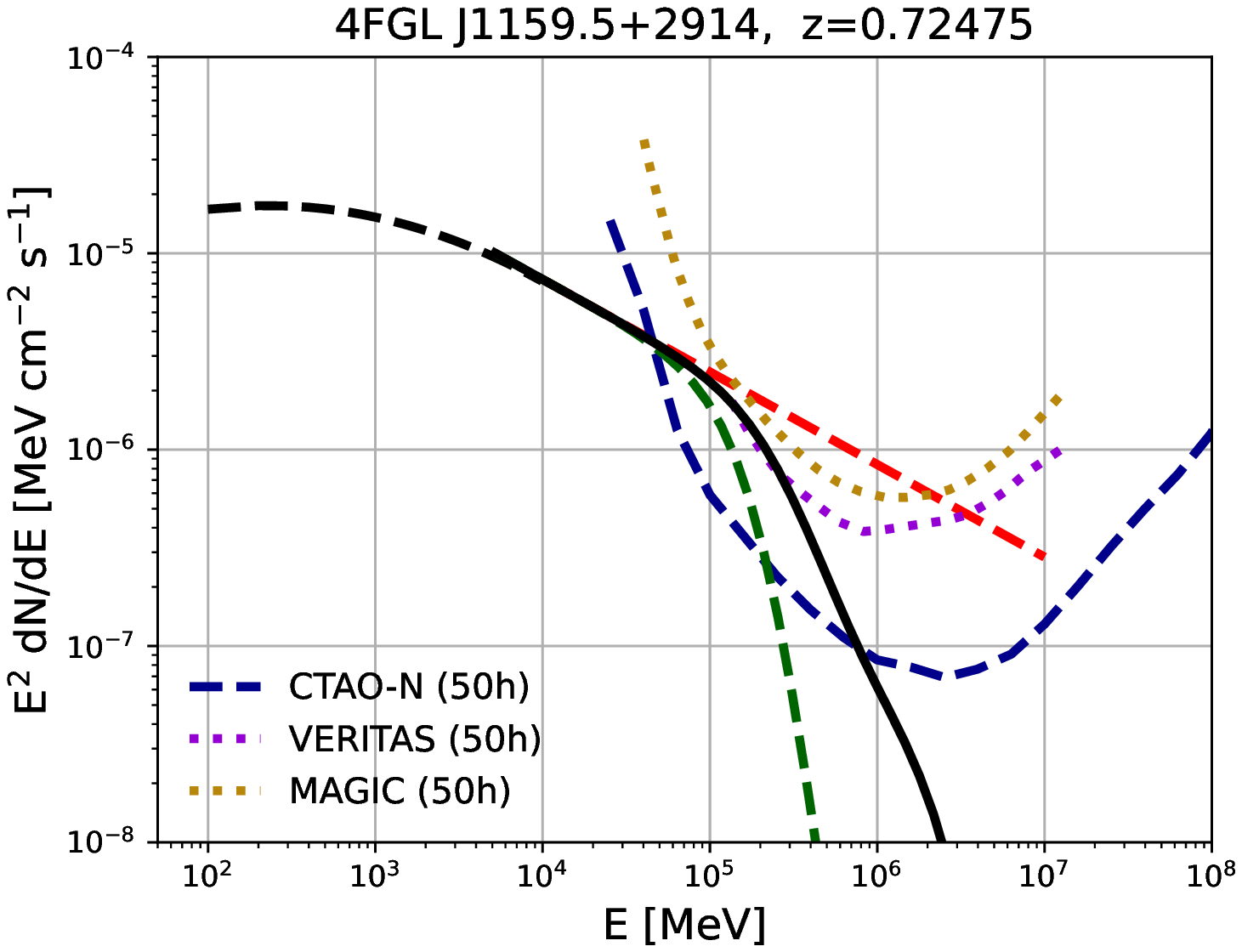}
\includegraphics[width=0.30\textwidth]{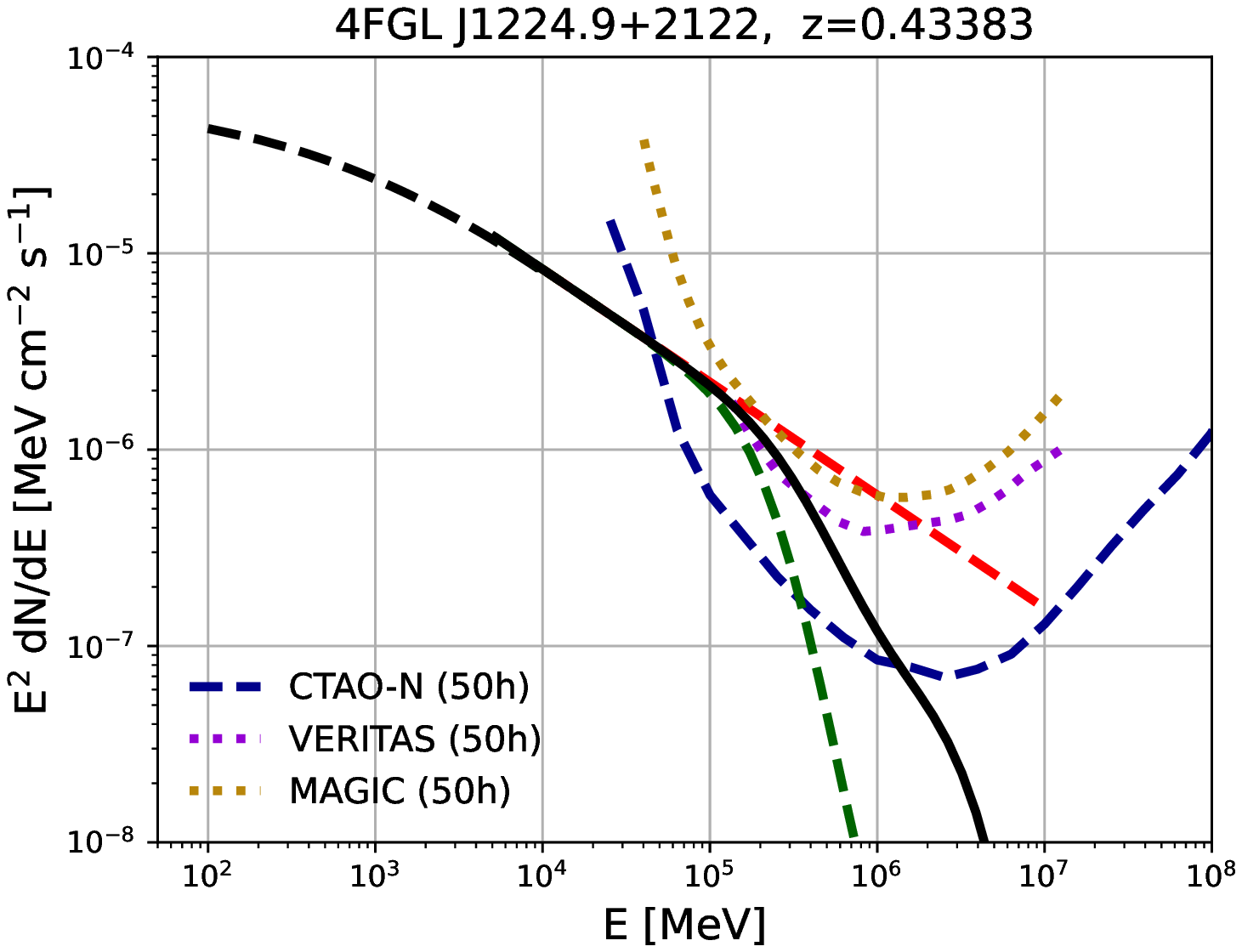}

\caption{32 VHE FSRQ candidates falling within the detection threshold of CTAO and other instruments. The black dashed line represents the \emph{Fermi} 4FGL-DR2 spectral fit. Red dashed line shows the linear extrapolation of \emph{Fermi} spectrum to VHE. The dashed green and solid black lines represent the EBL attenuated spectrum using \citet{Franceschini_2017} (FM) and modified EBL model (MFM) respectively. The differential sensitivities of \emph{CTAO}-North and South (Omega configuration, 50 hour), \emph{MAGIC} (50 hour), \emph{VERITAS} (50 hour) and H.E.S.S (50 hour) are also plotted. Sensitivity curves are given in different colors as depicted in the Figure labels.}
\label{fig:fig2}
\end{figure*}

\newpage
\begin{figure*}

\includegraphics[width=0.30\textwidth]{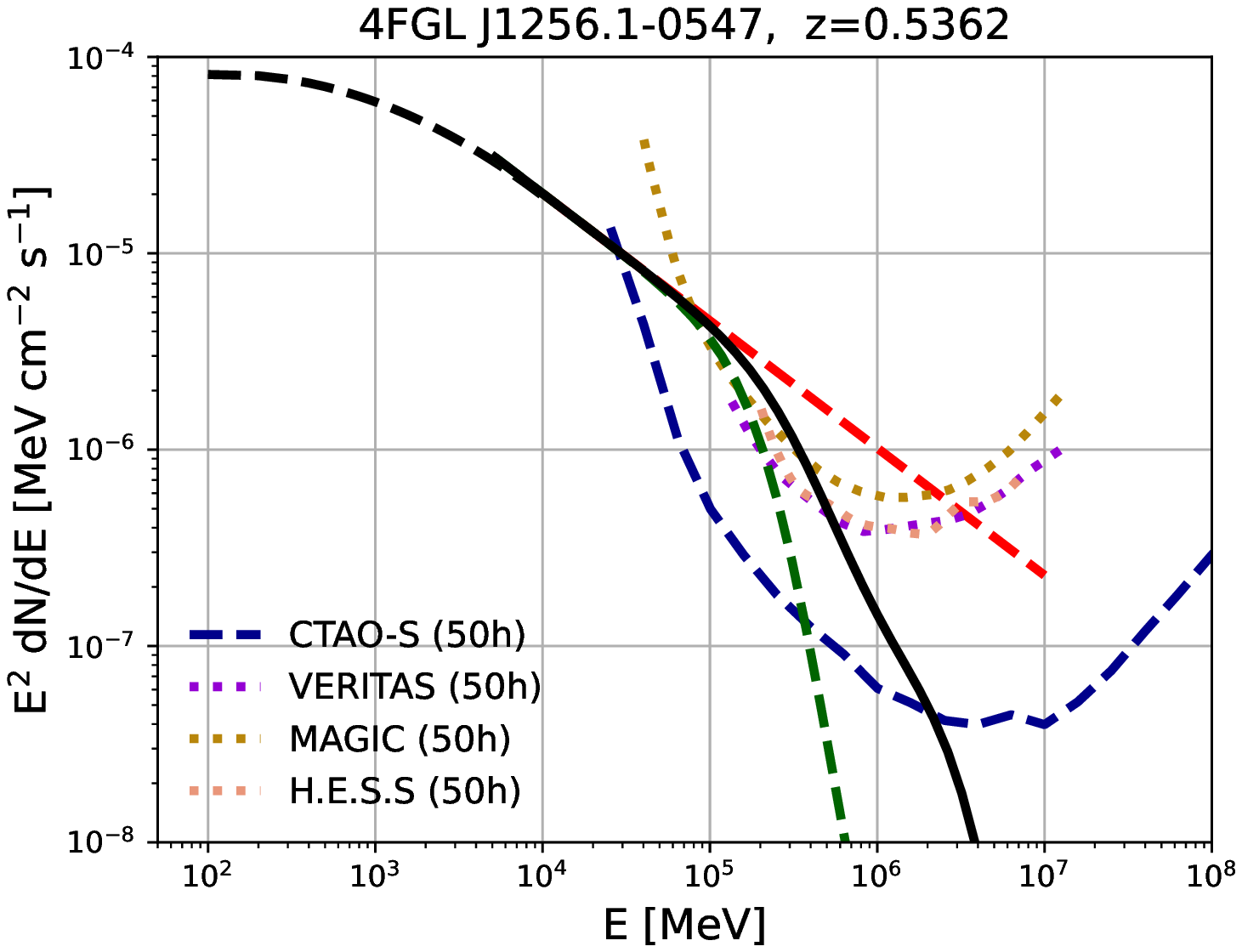}
\includegraphics[width=0.30\textwidth]{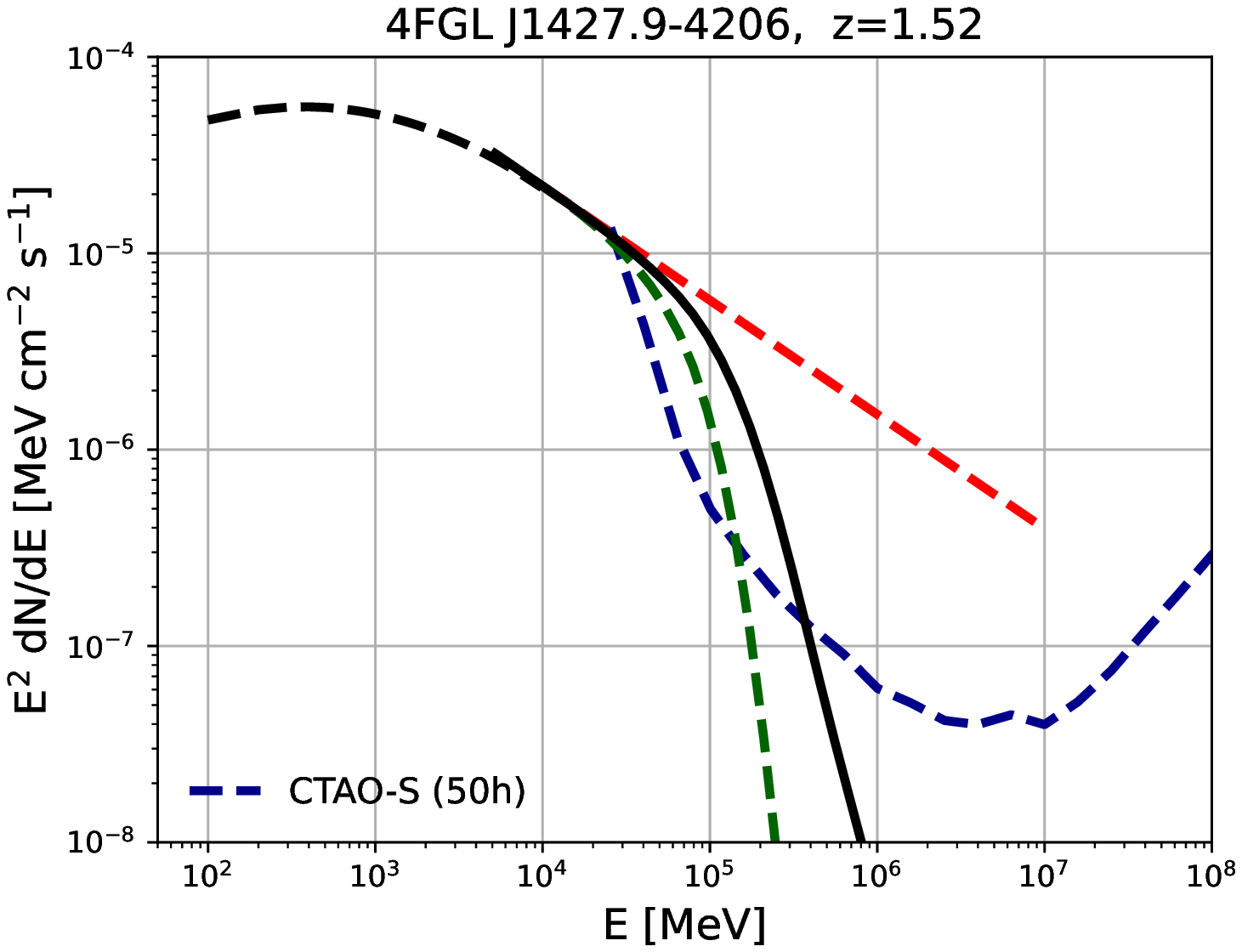}
\includegraphics[width=0.30\textwidth]{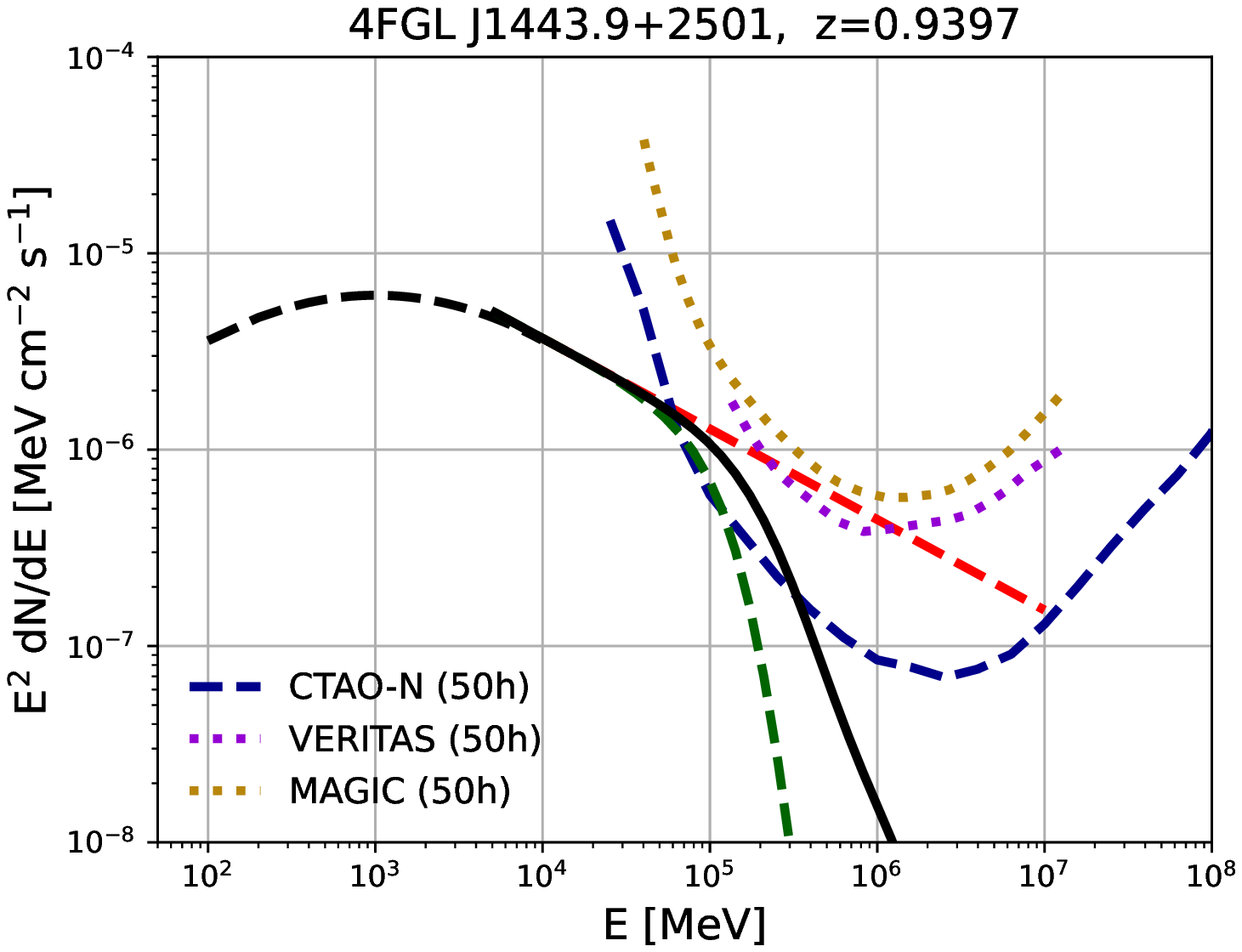}
\includegraphics[width=0.30\textwidth]{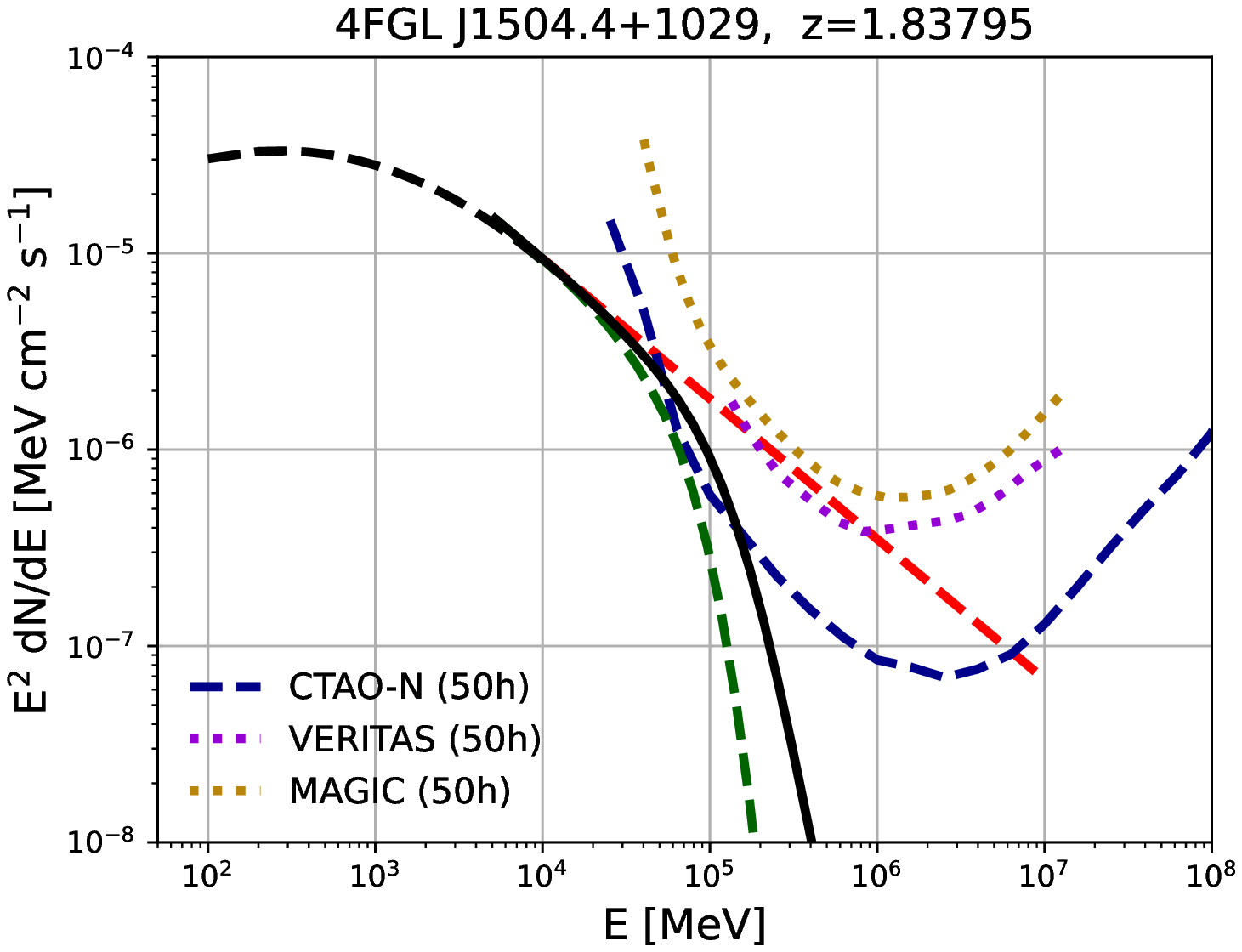}
\includegraphics[width=0.30\textwidth]{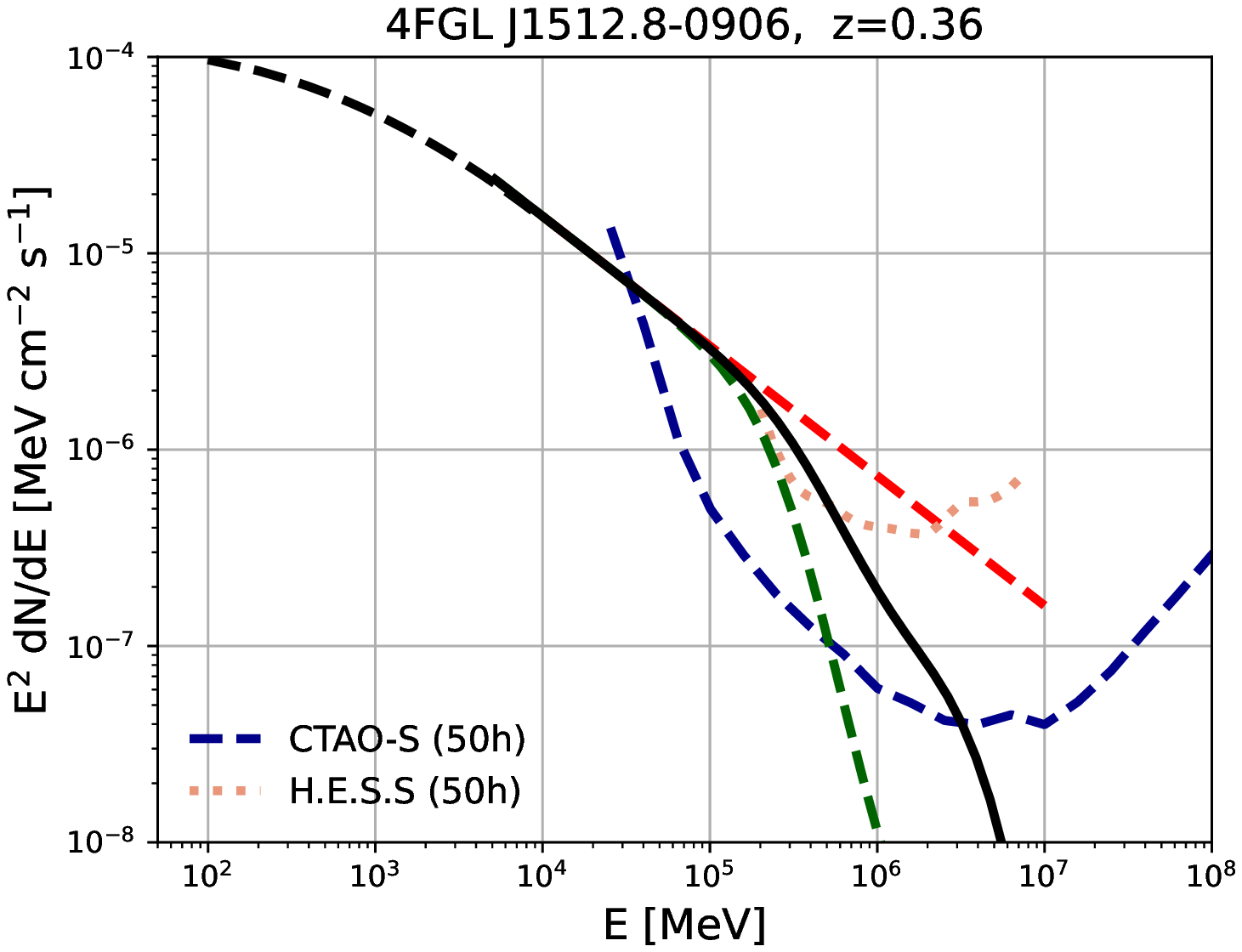}
\includegraphics[width=0.30\textwidth]{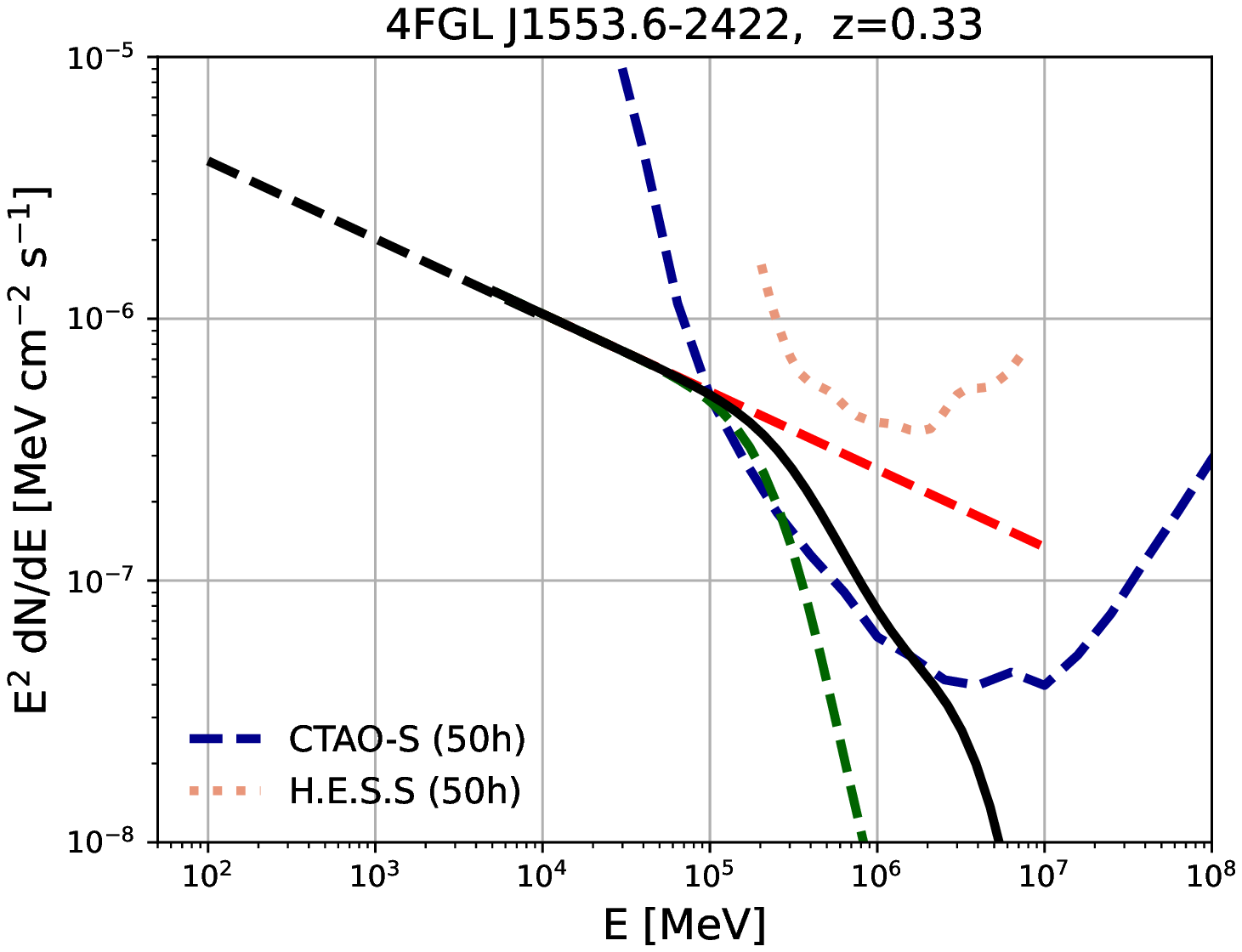}
\includegraphics[width=0.30\textwidth]{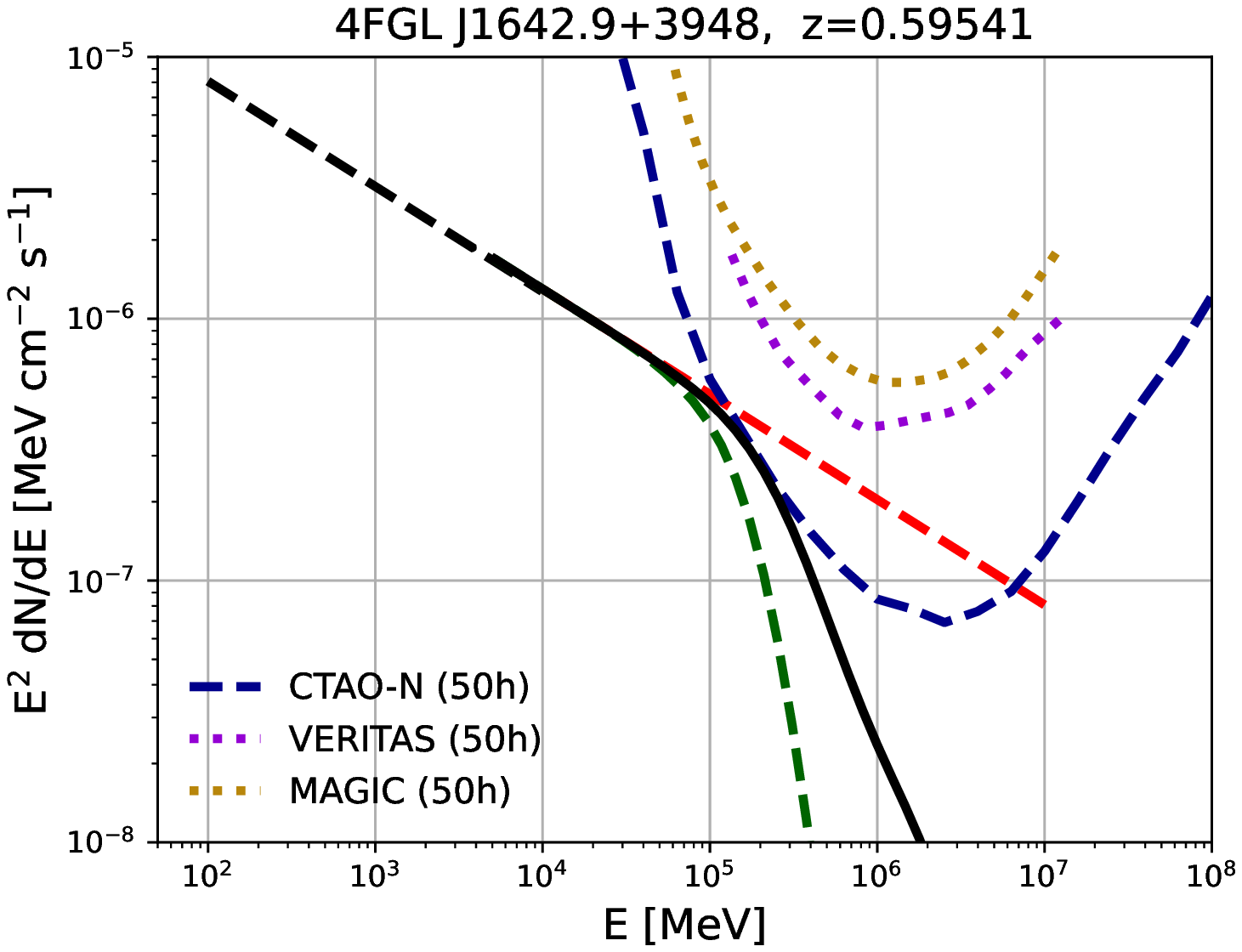}
\includegraphics[width=0.30\textwidth]{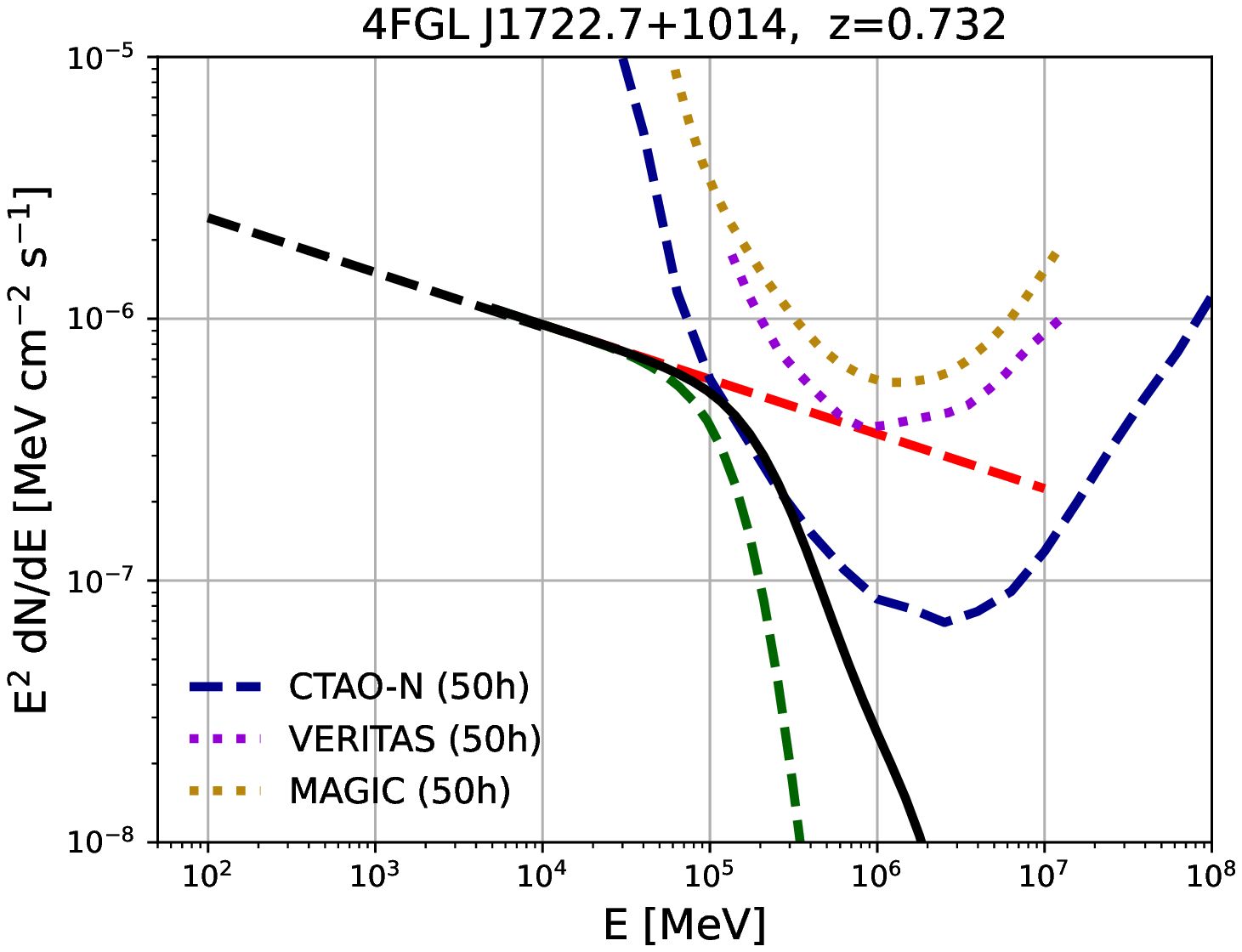}
\includegraphics[width=0.30\textwidth]{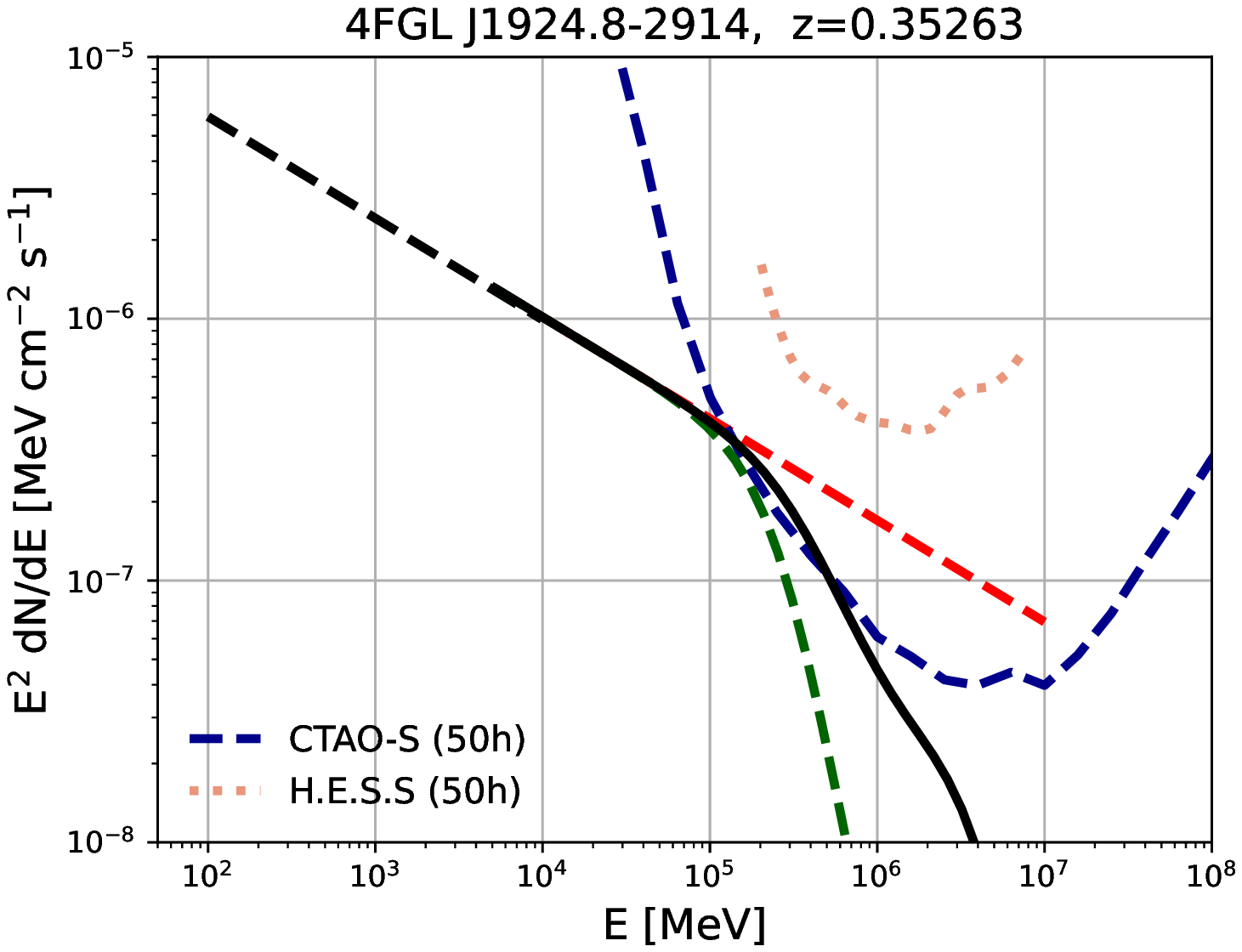}
\includegraphics[width=0.30\textwidth]{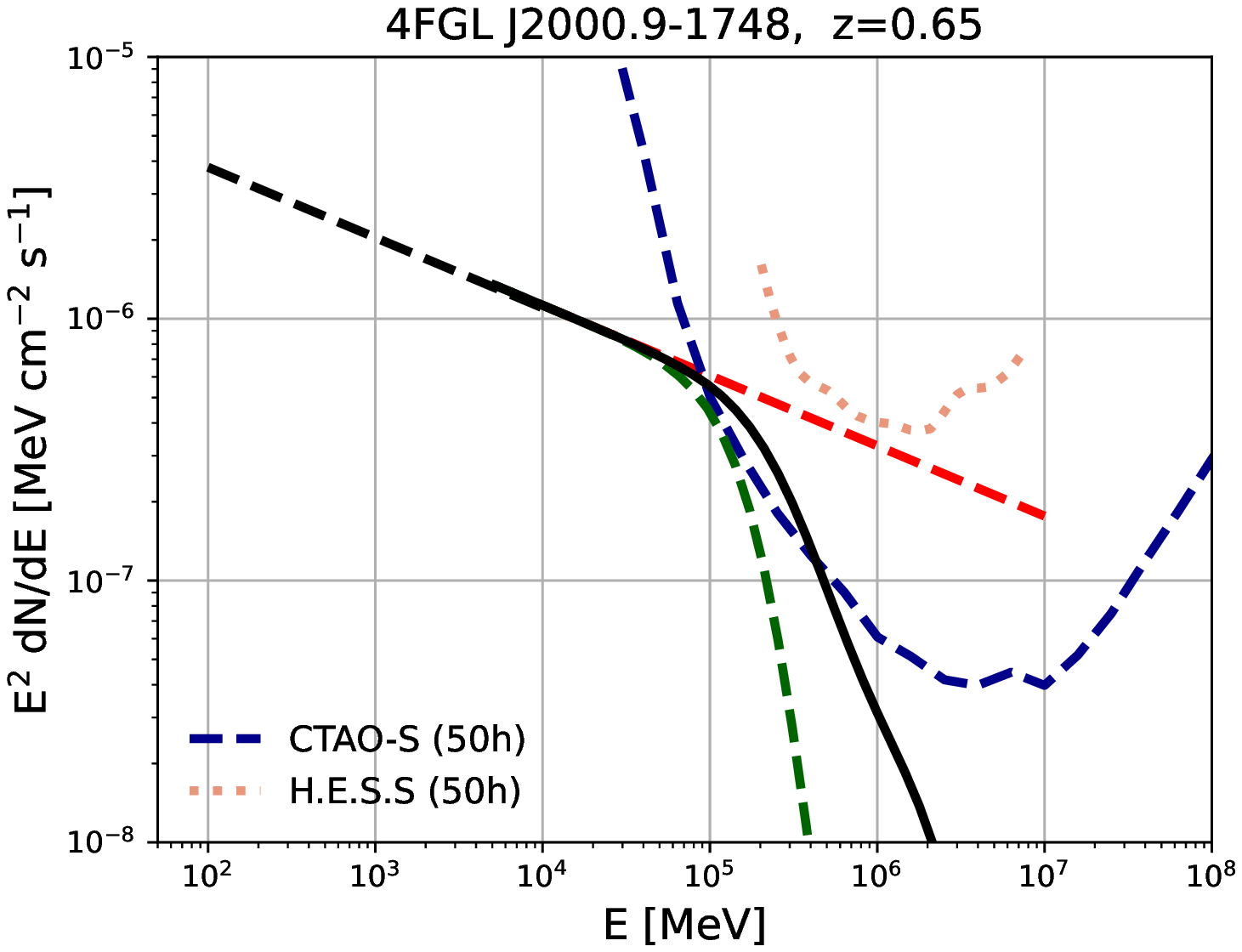}
\includegraphics[width=0.30\textwidth]{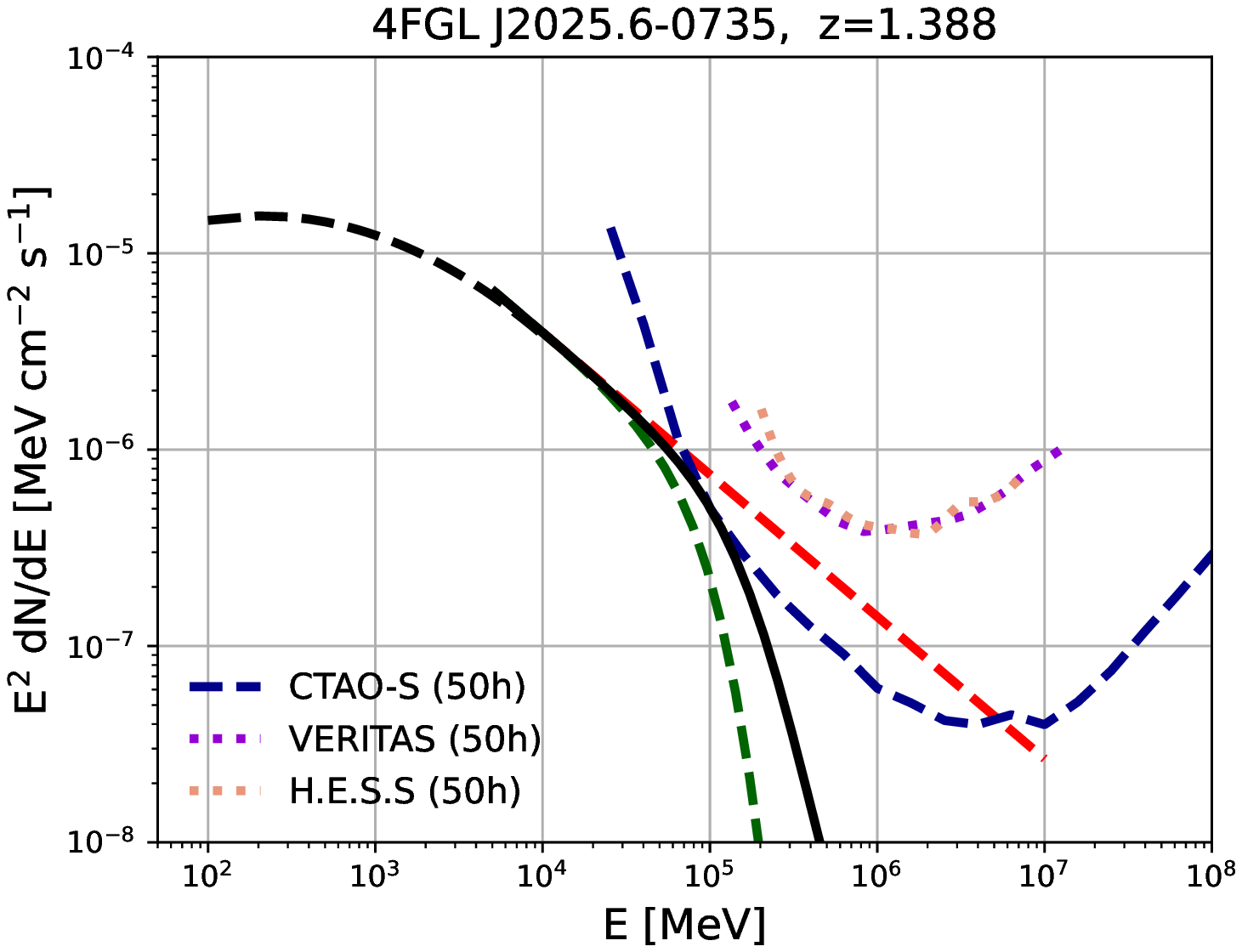}
\includegraphics[width=0.30\textwidth]{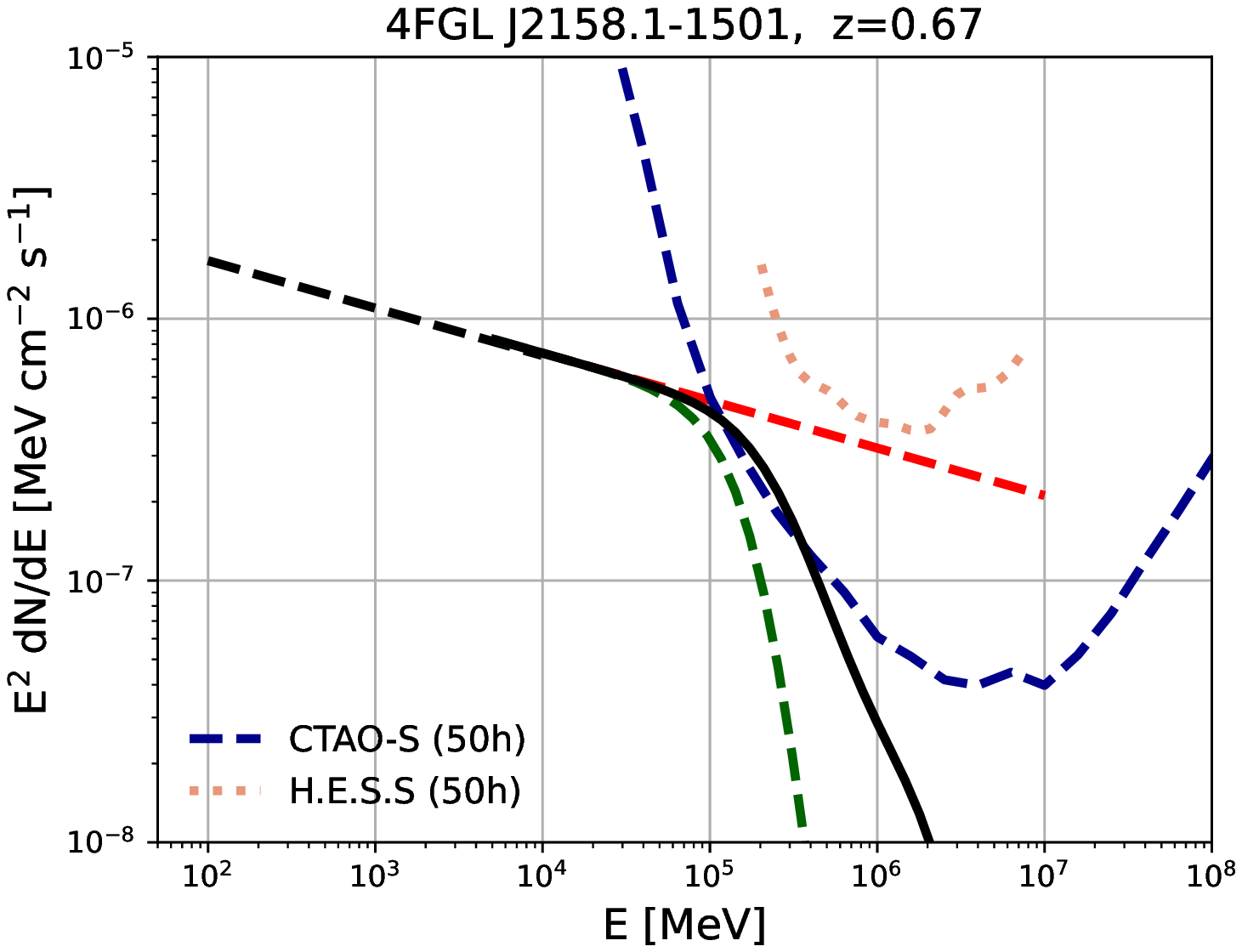}
\includegraphics[width=0.30\textwidth]{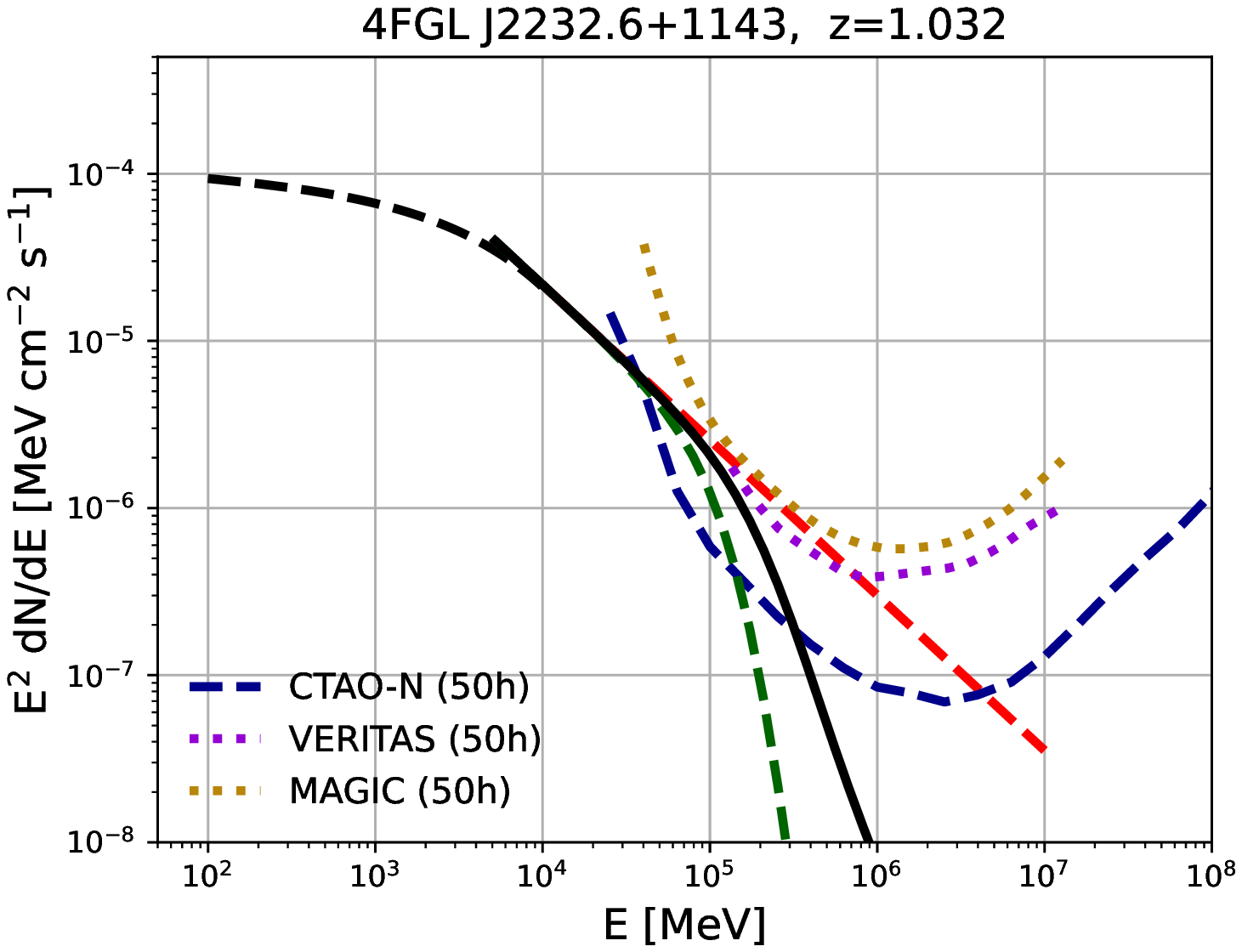}
\includegraphics[width=0.30\textwidth]{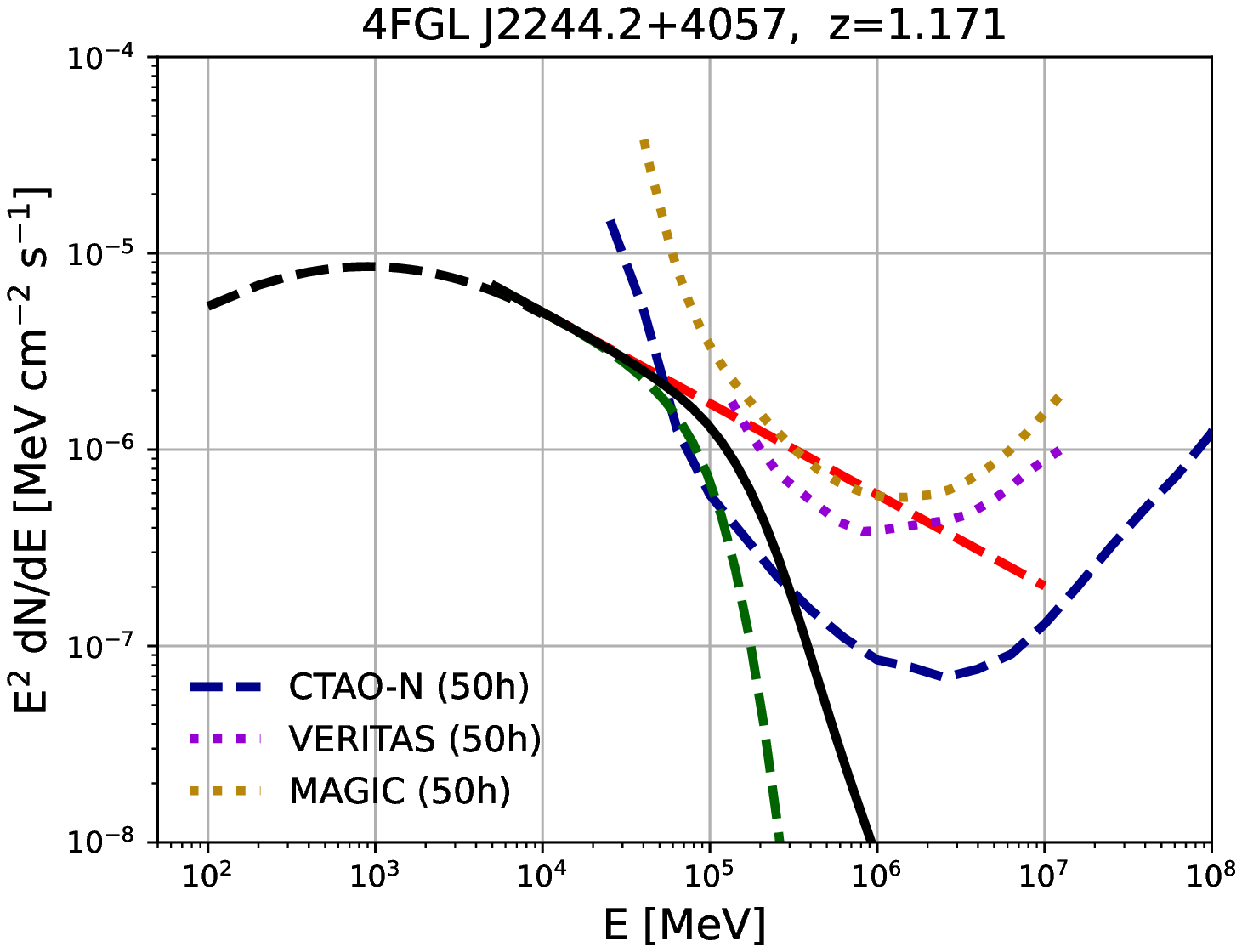}
\includegraphics[width=0.30\textwidth]{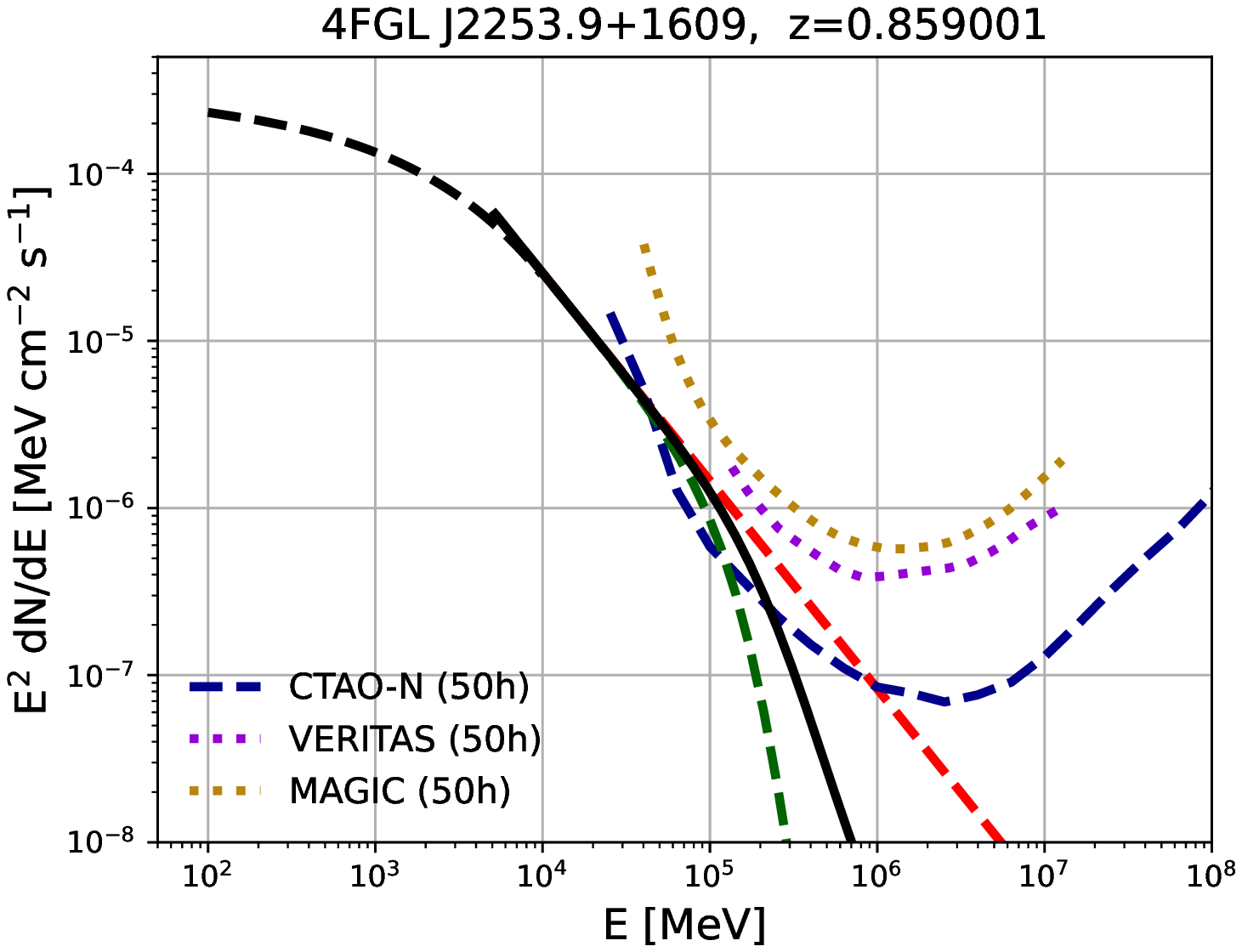}\\
\contcaption{}
\end{figure*}

\newpage
\begin{figure*}

\includegraphics[width=0.30\textwidth]{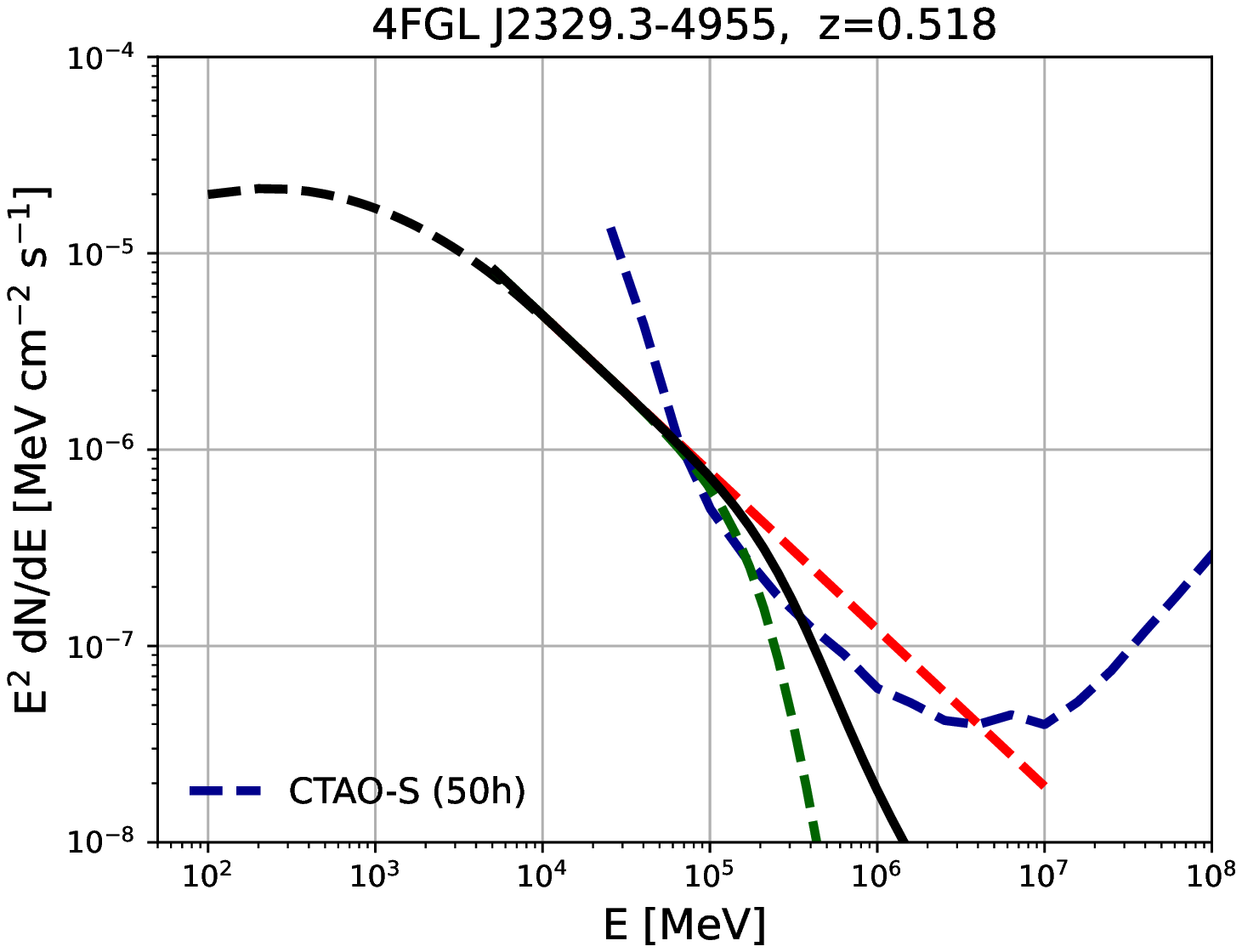}
\includegraphics[width=0.30\textwidth]{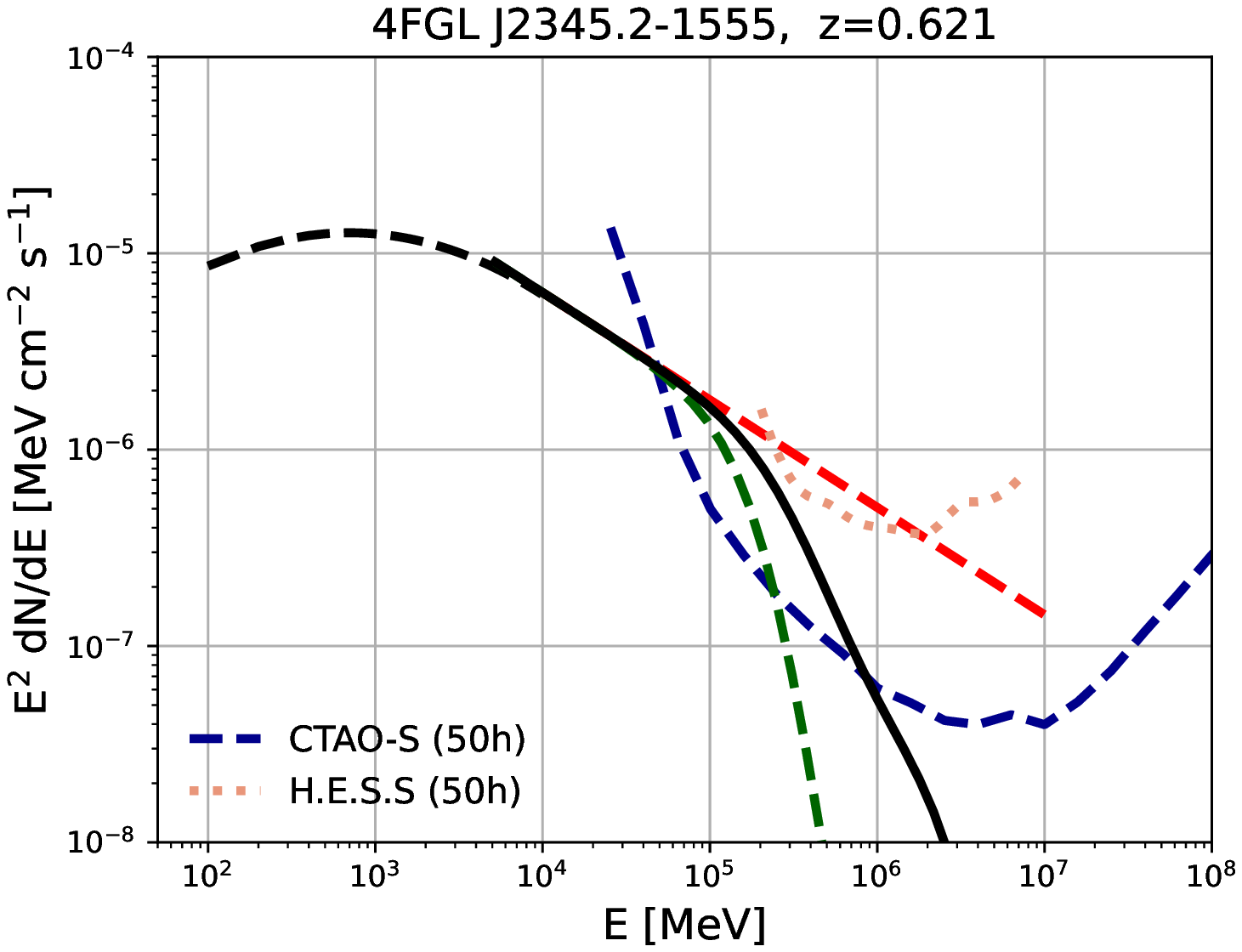}

\contcaption{}
\end{figure*}

\begin{landscape}
\begin{table}
\begin{center}
\caption{List of VHE FSRQ candidates falling within the detection threshold of CTAO (Omega and Alpha configurations) and other IACTs. D and N depict "detected" and "non detected" sources corresponding to different instruments within 50 hours of exposure time. The FM and MFM corresponds to the detection status using opacity estimates from \citet{Franceschini_2017} and modified EBL models respectively for each instrument.}
\label{tab:tab1}
\begin{tabular}{ccccccccccccccc}
\hline
\multirow{2}{*}{Sourcename (4FGL-DR2)}&\multirow{2}{*}{R.A.}&	\multirow{2}{*}{Decl.}&	\multirow{2}{*}{Redshift}& \multicolumn{1}{c}{Spectral Slope} & \multicolumn{2}{c}{CTAO (Omega)} & \multicolumn{2}{c}{CTA (Alpha)} & \multicolumn{2}{c}{VERITAS} & \multicolumn{2}{c}{MAGIC} & \multicolumn{2}{c}{HESS}\\\
&&&& (10 GeV) & MFM&FM & MFM&FM & MFM&FM  & MFM&FM & MFM&FM\\
\hline\hline
4FGL J0043.8+3425	&	00 43 53.2	&	+34 25 54	&	0.966	&	2.08	&	D&D	&	D&D	&	D&N	&	N&N	&		N&N\\
4FGL J0102.8+5824	&	01 02 48.2	&	+58 24 33	&	0.644	&	2.53	&	D&D	&	D&D	&	N&N	&	N&N	&		N&N\\
4FGL J0221.1+3556	&	02 21 07.4	&	+35 56 09	&	0.96	&	2.50	&	D&D	&	D&D	&	N&N	&	N&N	&		N&N\\
4FGL J0237.8+2848	&	02 37 53.7	&	+28 48 16	&	1.206	&	2.74	&	D&N	&	D&N	&	N&N	&	N&N	&		N&N\\
4FGL J0348.5-2749	&	03 48 34.0	&	-27 49 49	&	0.99	&	2.44	&	D&D	&	D&N	&	N&N	&	N&N	&		N&N\\
4FGL J0457.0-2324	&	04 57 02.6	&	-23 24 54	&	1.00	&	2.56	&	D&D	&	D&D	&	N&N	&	N&N	&		N&N\\
4FGL J0510.0+1800	&	05 10 04.3	&	+18 00 49	&	0.42	&	2.51	&	D&D	&	D&D	&	N&N	&	N&N	&		N&N\\
4FGL J0515.6-4556	&	05 15 37.4	&	-45 56 54	&	0.194	&	2.42	&	D&N	&	N&N	&	N&N	&	N&N	&		N&N\\
4FGL J0730.3-1141	&	07 30 18.6	&	-11 41 20	&	1.59	&	2.70	&	D&N	&	D&N	&	N&N	&	N&N	&		N&N\\
4FGL J0808.2-0751	&	08 08 15.6	&	-07 51 20	&	1.84	&	2.57	&	D&N	&	N&N	&	N&N	&	N&N	&		N&N\\
4FGL J0957.6+5523	&	09 57 39.9	&	+55 23 02	&	0.902000	&	2.30	&	D&D	&	D&D &	D&D	&	D&N	&	N&N\\
4FGL J1048.4+7143	&	10 48 25.6	&	+71 43 47	&	1.1500	&	2.66	&	D&N	&	D&N	&	N&N	&	N&N	&		N&N\\
4FGL J1127.0-1857	&	11 27 03.2	&	-18 57 50	&	1.05	&	2.56	&	D&D	&	D&D	&	N&N	&	N&N	&		N&N\\
4FGL J1159.5+2914	&	11 59 32.2	&	+29 14 41	&	0.72475	&	2.47	&	D&D	&	D&D	&	D&N	&	N&N	&		N&N\\
4FGL J1224.9+2122	&	12 24 54.6	&	+21 22 53	&	0.43383	&	2.57	&	D&D	&	D&D	&	D&N	&	N&N	&		N&N\\
4FGL J1256.1-0547	&	12 56 10.0	&	-05 47 19	&	0.53620	&	2.65	&	D&D	&	D&D	&	D&D	&	D&D	&		D&N\\
4FGL J1427.9-4206	&	14 27 56.8	&	-42 06 22	&	1.52	&	2.58	&	D&D	&	D&D	&	N&N	&	N&N	&		N&N\\
4FGL J1443.9+2501	&	14 43 58.4	&	+25 01 45	&	0.9397	&	2.46	&	D&D	&	D&D	&	N&N	&	N&N	&		N&N\\
4FGL J1504.4+1029	&	15 04 24.8	&	+10 29 52	&	1.83795	&	2.71	&	D&D	&	D&D	&	N&N	&	N&N	&		N&N\\
4FGL J1512.8-0906	&	15 12 51.5	&	-09 06 23	&	0.36	&	2.66	&	D&D	&	D&D	&	N&N	&	N&N	&		D&D\\
4FGL J1553.6-2422	&	15 53 36.6	&	-24 22 07	&	0.33	&	2.30	&	D&D	&	D&D	&	N&N	&	N&N	&		N&N\\
4FGL J1642.9+3948	&	16 42 56.2	&	+39 48 59	&	0.59541	&	2.40	&	D&N	&	D&N	&	N&N &	N&N &  N&N\\
4FGL J1722.7+1014	&	17 22 44.6	&	+10 14 05	&	0.732	&	2.20	&	D&N	&	D&N	&	N&N	&	N&N	&		N&N\\
4FGL J1924.8-2914	&	19 24 51.3	&	-29 14 48	&	0.35263	&	2.38 &	D&D	&	D&N	&	N&N	&	N&N	&		N&N\\
4FGL J2000.9-1748	&	20 00 56.3	&	-17 48 59	&	0.65	&	2.27	&	D&D	&	D&N	&	N&N	&	N&N	&		N&N\\
4FGL J2025.6-0735	&	20 25 41.3	&	-07 35 40	&	1.388000	&	2.72	&	D&N	&	D&N	&	N&N	&	N&N	& N&N\\
4FGL J2158.1-1501	&	21 58 06.6	&	-15 01 25	&	0.67	&	2.18	&	D&N	&	D&N	&	N&N	&	N&N	&		N&N\\
4FGL J2232.6+1143	&	22 32 36.6	&	+11 43 50	&	1.032	&	2.93	&	D&D	&	D&D	&	N&N	&	N&N	&		N&N\\
4FGL J2244.2+4057	&	22 44 14.7	&	+40 57 35	&	1.171	&	2.46	&	D&D	&	D&D	&	N&N	&	N&N	&		N&N\\
4FGL J2253.9+1609	&	22 53 59.1	&	+16 09 02	&	0.859001	&	3.24	&	D&D	&	D&D	&	N&N	&	N&N	&		N&N\\
4FGL J2329.3-4955	&	23 29 19.1	&	-49 55 57	&	0.518	&	2.80	&	D&D	&	D&D	&	N&N	&	N&N	&		N&N\\
4FGL J2345.2-1555	&	23 45 12.7	&	-15 55 06	& 0.621 & 2.55 &	D&D	&	D&D	&	N&N	&	N&N	&		N&N\\

\hline
\end{tabular}
\end{center}
\end{table}
\end{landscape}

\newpage
\begin{landscape}
\begin{table}
\begin{center}
\caption{List of VHE FSRQ candidates falling within the marginal detection threshold of CTAO (Omega and Alpha configurations). M and N depict "marginally detected" and "non detected" sources corresponding to \emph{CTAO} within 5 hours of exposure time. The FM and MFM corresponds to the detection status using opacity estimates from \citet{Franceschini_2017} and modified EBL models respectively.}
\label{tab:tab2}
\begin{tabular}{ccccccccc}
\hline
\multirow{2}{*}{Sourcename (4FGL-DR2)}&\multirow{2}{*}{R.A.}&	\multirow{2}{*}{Decl.}&	\multirow{2}{*}{Redshift}& \multicolumn{1}{c}{Spectral Slope}&\multicolumn{2}{c}{CTAO (Omega)} & \multicolumn{2}{c}{CTAO (Alpha)}\\
&&&& (10 GeV) & MFM&FM & MFM&FM\\
\hline\hline
4FGL J0038.2-2459	&	00 38 15.6	&	-24 59 24	&	0.49806	&	2.64	&	M&N	&	M & N\\
4FGL J0050.4-0452	&	00 50 26.9	&	-04 52 50	&	0.922	&	2.38	&	M&N	&	M&N\\
4FGL J0108.6+0134	&	01 08 40.7	&	+01 34 55	&	2.108898	&	2.89	&	M&M	&	M&M\\
4FGL J0112.8+3208	&	01 12 53.4  &   +32 08 24	&	0.6100	&	2.65	&	M&M	&	M&M\\
4FGL J0113.4+4948	&	01 13 28.4 & +49 48 19	&	0.39	&	2.45	&	M&M	&	M&M\\
4FGL J0116.0-1136 	&	01 16 00.1&-11 36 22	&	0.671	&	2.36	&	M&M	&	M&M\\
4FGL J0118.9-2141	&	01 18 54.1&-21 41 41	&	1.16	&	2.65	&	M&M	&	M&M\\
4FGL J0128.5+4440	&	01 28 34.5&+44 40 40	&	0.228	&	2.13	&	M&N	&	N&N\\
4FGL J0132.7-1654	&	01 32 42.2&-16 54 37	&	1.02	&	2.64	&	M&N	&	M&N\\
4FGL J0133.1-5201	&	01 33 10.5&-52 01 13	&	0.02	&	2.69	&   M&M	&	M&M\\
4FGL J0137.0+4751	&	01 37 02.5&+47 51 49	&	0.86	&	2.72	&	M&M	&	M&M\\
4FGL J0206.4-1151	&	02 06 24.4&-11 51 27	&	1.663	&	2.40	&	M&N	&	N&N\\
4FGL J0210.7-5101	& 02 10 46.7&-51 01 18	&	1.003	&	2.76	&   M&M	&	M&M\\
4FGL J0217.8+0144	& 02 17 50.9&+01 44 05		&	1.72	&	2.51	&	M&N	&	M&N\\
4FGL J0236.8-6136	& 02 36 48.5&-61 36 38	&	0.466569	&	2.35	&	M&M	&	M&M\\
4FGL J0252.8-2219	& 02 52 48.2&-22 19 13		&	1.419	&	2.80	&	M&N	&	M&N\\
4FGL J0253.5+3216	&	02 53 31.9&+32 16 57	&	0.859	&	2.17	&	M&N	&	M&N\\
4FGL J0259.4+0746	&	02 59 25.9&+07 47 00	&	0.89	&	2.41	&	M&M	&	M&M\\
4FGL J0312.8+0134	& 03 12 53.3&+01 34 21		&	0.664	&	2.33	&	M&N	&	N&N\\
4FGL J0339.5-0146	& 03 39 30.5&-01 46 37	&	0.85	&	2.72 &	M&M	&	M&M\\
4FGL J0423.3-0120	& 04 23 18.2&-01 20 03		&	0.91609	&	2.73	&	M&M	&	M&M\\
4FGL J0442.6-0017	& 04 42 38.7&-00 17 46		&	0.845	& 2.78	&	M&N	&	M&N\\
4FGL J0449.2+6329	& 04 49 16.4&+63 29 40		&	0.781	&	2.60	&	M&N	&	M&N\\
4FGL J0505.3+0459	& 05 05 22.3&+04 59 58		&	0.59	&	2.85	&	M&M	&	M&M\\
4FGL J0509.4+1012	&  05 09 24.2&+10 12 03		&	0.621	&	2.41	&	M&N	&	M&N\\
4FGL J0526.2-4830	& 05 26 17.1&-48 30 54		&	1.3041	&	2.55	&	M&M	&	M&M\\
4FGL J0532.6+0732	&	05 32 41.3&+07 32 57	&	1.254	&	2.78	&	M&M	&	M&M\\
4FGL J0532.9-8325	&	05 32 58.9&-83 25 57	&	0.774	&	2.07	&	M&N	&	M&N\\
4FGL J0654.4+4514	& 06 54 25.4&+45 14 41	&	0.928	&	2.57	&	M&N	&	M&N\\
4FGL J0709.7-0255	& 07 09 46.8&-02 55 48	&	1.472	&	2.54	&	M&N	&	M&N\\
4FGL J0719.3+3307	& 07 19 21.6&+33 07 24		&	0.779	&	2.63	&	M&M	&	M&M\\
4FGL J0725.2+1425	&	07 25 17.8&+14 25 16	&	1.038	&	2.58	&	M&M	&	M&M\\
4FGL J0739.2+0137	&	07 39 16.8&+01 37 18	&	0.19	&	2.89	&   M&M	&	M&M\\
4FGL J0742.6+5443	&	07 42 41.2&+54 43 37	&	0.723 	&	2.70	&	M&N	&	M&N\\
4FGL J0748.6+2400	& 07 48 39.3&+24 01 00		& 0.40932	    &	2.34	&	M&M	&	M&M\\
4FGL J0829.4+0857	& 08 29 24.8&+08 57 22	&	0.866	&	2.06	&	M&N	&	M&N\\
4FGL J0850.1-1212	& 08 50 09.9&-12 12 44		&	0.87	&	2.69	&	M&M	&	M&M\\
4FGL J0909.7-0230	& 09 09 47.1&-02 30 51		&	0.957000	&	2.73	&	M&N	&	N&N\\
4FGL J0921.6+6216	& 09 21 40.4&+62 16 15		&	1.447000	&	2.69	&	M&N	&	M&N\\
4FGL J0922.4-0528	&	09 22 27.0&-05 28 29	&	0.974	&	2.14	&	M&M	&	M&N\\
4FGL J1006.7-2159	&	10 06 46.3&-21 59 28	&	0.33	&	2.57	&	M&M	&	M&M\\
4FGL J1016.0+0512	& 10 16 02.2&+05 12 32		&	1.701000	&	2.22	&	M&N	&	M&N\\
4FGL J1033.1+4115	& 10 33 06.0&+41 15 43		&	1.118000	&	2.41	&	M&N	&	M&N\\
4FGL J1033.9+6050	& 10 33 56.4&+60 50 57	&	1.408000	&	2.56	&	M&M	&	M&M\\
4FGL J1037.4-2933	&	10 37 25.5&-29 33 24	&	0.31	&	2.41	&	M&N	&	N&N\\
4FGL J1043.2+2408	& 10 43 13.3&+24 08 46		&	0.560000	& 2.32	&	M&M	&	M&M\\
\hline
\end{tabular}
\end{center}
\end{table}
\end{landscape}

\newpage
\begin{landscape}
\begin{table}
\begin{center}
\contcaption{}
\begin{tabular}{ccccccccc}
\hline
\multirow{2}{*}{Sourcename (4FGL-DR2)}&\multirow{2}{*}{R.A.}&	\multirow{2}{*}{Decl.}&	\multirow{2}{*}{Redshift}& \multicolumn{1}{c}{Spectral Slope}&\multicolumn{2}{c}{CTAO (Omega)} & \multicolumn{2}{c}{CTAO (Alpha)}\\
&&&& (10 GeV) & MFM&FM & MFM&FM\\
\hline\hline
4FGL J1106.0+2813	& 11 06 00.5&+28 13 31	&	0.84434	&	2.37	&	M&N	&	M&N\\
4FGL J1123.4-2529	& 11 23 28.4&-25 29 17		&	0.146	&	2.24	&	M&M	&	M&N\\
4FGL J1127.8+3618	& 11 27 51.3&+36 18 50		&	0.8841	&	2.37	&	M&N	&	M&N\\
4FGL J1146.9+3958	& 11 46 57.7&+39 58 39		&	1.087885	&	2.71	&	M&M	&	M&M\\
4FGL J1153.4+4931	& 11 53 24.1&+49 31 01		&	0.33364	&	2.41	&	M&M	&	M&M\\
4FGL J1154.0+4037	&	11 54 03.5&+40 37 55	&	0.92834	&	2.11	&	M&N	&	M&N\\
4FGL J1246.7-2548	& 12 46 45.3&-25 48 06	&	0.63	&	2.85	&	M&M	&	M&M\\
4FGL J1310.5+3221	& 13 10 31.8&+32 21 17		&	0.99725	&	2.59	&	M&M	&	M&M\\
4FGL J1316.1-3338	& 13 16 06.0&-33 38 11		&	1.21	&	2.62	&	M&N	&	M&N\\
4FGL J1322.2+0842	& 13 22 12.2&+08 42 13		&	0.326	&	2.26	&	M&N	&	N&N\\
4FGL J1345.5+4453	& 13 45 34.6&+44 53 04		&	2.542000	&	2.65	&	M&M	&	M&M\\
4FGL J1349.5-1131	&	13 49 32.9&-11 31 08	&	0.340000	&	2.49	&	M&N	&	M&N\\
4FGL J1401.2-0915	& 14 01 13.5&-09 15 28		&	0.667	&	2.08	&	M&N	&	M&N\\
4FGL J1419.4-0838	& 14 19 26.4&-08 38 30		&	0.903	&	2.56	&	M&M	&	M&M\\
4FGL J1459.5+1527	&	14 59 33.0&+15 27 51	&	0.370	&	2.10	&	M&N	&	M&N\\
4FGL J1512.2+0202	& 15 12 16.8&+02 02 25	&	0.21945	&	2.68	&	M&M	&	M&M\\
4FGL J1522.1+3144	& 15 22 10.9&+31 44 22		&	1.4886	&	2.79	&	M&M	&	M&M\\
4FGL J1549.5+0236	&	15 49 32.4&+02 36 30	&	0.41421	&	2.41	&	M&M	&	M&M\\
4FGL J1625.7-2527	& 16 25 46.9&-25 27 54		&	0.79	&	2.90	&	M&M	&	M&M\\
4FGL J1635.2+3808	&	16 35 16.0&+38 08 24	&	1.814000	&	3.07	&	M&N	&	M&N\\
4FGL J1637.7+4717	& 16 37 44.2&+47 17 29		&	0.735000	&	2.43	&	M&N	&	M&N\\
4FGL J1640.4+3945	& 16 40 28.6&+39 45 45		&	1.672000	& 2.42	&	M&N	&	M&N\\
4FGL J1728.0+1216	& 17 28 04.8&+12 16 32		&	0.586	&	2.45	&	M&N	&	N&N\\
4FGL J1733.0-1305	& 17 33 03.2&-13 05 09		&	0.90	&	2.86	&   M&M	&	M&M\\
4FGL J1734.3+3858	& 17 34 23.6&+38 58 35		&	0.975	&	2.67	&	M&N	&	M&N\\
4FGL J1740.0+4737	& 17 40 05.4&+47 37 11		&	0.95	&	2.10	&	M&N	&	M&N\\
4FGL J1740.5+5211	& 17 40 32.6&+52 11 34		&	1.383	&	2.47	&	M&N	&	M&N\\
4FGL J1802.6-3940	& 18 02 41.1&-39 40 07		&	1.319	&	2.90	&	M&N	&	M&N\\
4FGL J1830.1+0617	& 18 30 08.7&+06 17 16		&	0.745	&	2.37	&	M&N	&	M&N\\
4FGL J1833.6-2103	& 18 33 38.4&-21 03 27		&	2.507	&	2.99	&	M&N	&	M&N\\
4FGL J1849.2+6705	& 18 49 16.6&+67 05 27		&	0.66	&	2.71	&	M&M	&	M&M\\
4FGL J1852.4+4856	& 18 52 27.8&+48 56 06		&	1.250	&	2.29	&	M&M	&	M&M\\
4FGL J2023.6-1139	& 20 23 36.8&-11 39 31	&	0.698	&	2.51	&	M&M	&	M&M\\
4FGL J2025.2+0317	& 20 25 14.0&+03 17 21	&	2.210	&	2.18	&	M&N	&	M&N\\
4FGL J2121.0+1901	& 21 21 02.4&+19 01 57		&	2.180	&	2.43	&	M&N	&	M&N\\
4FGL J2143.5+1743	&	21 43 34.6&+17 43 50	&	0.21	&	2.69	&	M&M	&	M&M\\
4FGL J2147.3-7536	& 21 47 18.4&-75 36 09	&	1.14	&	2.77	&	M&N	&	M&N\\
4FGL J2156.3-0036	& 21 56 19.1&-00 36 14		&	0.495	&	2.25	&	M&M	&	M&M\\
4FGL J2203.4+1725	& 22 03 29.3&+17 25 54		&	1.076	&	2.65	&	M&N	&	M&N\\
4FGL J2225.7-0457	& 22 25 43.7&-04 57 13		&	1.404	&	2.42	&	M&N	&	N&N\\
4FGL J2321.9+2734	& 23 21 58.1&+27 34 18		&	1.25	&	2.15	&	M&N	&	M&N\\
4FGL J2321.9+3204	& 23 21 54.7&+32 04 25		&	1.489	&	2.76	&	M&N	&	M&N\\
4FGL J2323.5-0317	& 23 23 33.2&-03 17 24		&	1.41	&	2.69	&	M&M	&	M&N\\
4FGL J2348.0-1630	& 23 48 03.8&-16 30 58		&	0.58	&	2.57	&	M&M	&	M&M\\

\hline
\end{tabular}
\end{center}
\end{table}
\end{landscape}

\bibliographystyle{mnras}
\bibliography{biblography} 


\bsp	
\label{lastpage}
\end{document}